\documentclass[12pt]{article}

\pdfoutput=1
\usepackage[latin9]{inputenc}
\usepackage[top=80pt,bottom=85pt,left=67pt,right=67pt]{geometry}
\usepackage{amssymb,amsmath}
\usepackage{xypic,subfigure}
\usepackage{graphicx,color}
\usepackage{bbm}
\usepackage{float}
\usepackage{cite}
\usepackage{tensor}
\usepackage{slashed}
\usepackage[debug,pageanchor=false]{hyperref}
\definecolor{link}{rgb}{.8,.15,.1}
\hypersetup{colorlinks=true,linkcolor=link,citecolor=link,urlcolor=link,linktocpage}

 \allowdisplaybreaks

\makeatletter
\@addtoreset{equation}{section}
\makeatother

\newcommand{\beq}{\begin{equation}}
\newcommand{\eeq}{\end{equation}}
\newcommand{\bea}{\begin{eqnarray}}
\newcommand{\eea}{\end{eqnarray}}
\newcommand{\nn}{\nonumber}

\newcommand{\eq}{\begin{equation}}
\newcommand{\feq}{\end{equation}}
\newcommand{\eqn}{\begin{eqnarray}}
\newcommand{\feqn}{\end{eqnarray}}

\begin{document}

\begin{titlepage}

\begin{center}

\vskip .5in 
\noindent

{\Large \bf{${\cal N}=(4,4)$ supersymmetric AdS$_3$ solutions in $d=11$}}

\bigskip\medskip

Andrea Conti\footnote{contiandrea@uniovi.es},  Niall T. Macpherson\footnote{macphersonniall@uniovi.es}  \\

\bigskip\medskip
{\small 

Department of Physics, University of Oviedo,
Avda. Federico Garcia Lorca s/n, 33007 Oviedo}

\medskip
{\small and}

\medskip
{\small 

Instituto Universitario de Ciencias y Tecnolog\'ias Espaciales de Asturias (ICTEA),\\
Calle de la Independencia 13, 33004 Oviedo, Spain}

\vskip 2cm 

     	{\bf Abstract }\\[2mm]
			We derive necessary and sufficient conditions for AdS$_3$ solutions of $d=11$ supergravity to preserve  ${\cal N}=(1,1)$ supersymmetry in terms of G-structures. Such solutions necessarily support an SU(3)-structure on the internal 8-manifold M$_8$, in terms of which we phrase the conditions for supersymmetry preservation. We use this to derive the local form of all ${\cal N}=(4,4)$ supersymmetric AdS$_3$ solutions in $d=11$, for which M$_8$ decomposes as a foliation of a 3-sphere over a 5 dimensional base. There are 3 independent classes, 2 of which preserve the small superconformal algebra and one preserving its large counterpart for which M$_5$ contains a second 3-sphere. We show that for each solution with large (4,4) supersymmetry there are two corresponding solutions with small $(4,4)$, one for which M$_5$  maintains its 3-sphere, one where this blows up to $\mathbb{R}^3$ which can be compactified to $\mathbb{T}^3$. We use our results to construct several new solutions that lie within our derived classes as well as recovering some existing solutions.
     	\end{center}
     	\noindent

\noindent

\vfill
\eject

\end{titlepage}

\tableofcontents

\newpage

\section{Introduction}

In recent years there has been significant effort made to classify and construct supersymmetric AdS$_3$ solutions of ten and eleven dimensional supergravity (see for instance \cite{Couzens:2017way,Couzens:2017nnr,Dibitetto:2018ftj,Dibitetto:2018iar,Macpherson:2018mif,Lozano:2019emq,Couzens:2019iog,Legramandi:2019xqd,Lozano:2020bxo,Faedo:2020lyw,Faedo:2020nol,Couzens:2021tnv,Eloy:2021fhc,Couzens:2021veb,Emelin:2022cac,Ashmore:2022ydf,Anabalon:2022fti,VanHemelryck:2022ynr,Lima:2022hji,Macpherson:2022sbs,Eloy:2023zzh,Eloy:2023acy,Farakos:2023wps,Farakos:2023nms,Itsios:2023kma,Arboleya:2024vnp,Eloy:2024lwn,Lozano:2024idt,Conti:2024rwd,Conti:2024qgx}). In large part this has been  motivated by the following: First there is the appearance of AdS$_3$ factors in the near horizon limit of black string solutions making such constructions relevant to the counting of microstates. Second is that AdS$_3$ solutions are important to the AdS/CFT correspondence, whose  AdS$_3$/CFT$_2$ avatar is the most under control from both the string and CFT perspective. There is also the fact that AdS$_3$ factors appear in solutions dual to defects within higher dimensional CFTs and interfaces.

Supersymmetric AdS$_3$ solutions and their dual SCFTs stand out when compared to their higher dimensional counterparts because they allow for many more district ways to realise  extended  superconformal symmetry \cite{Fradkin:1992bz,Beck:2017wpm}. This follows first from the fact that the relevant superconformal algebras decompose as a direct sum of sub algebras of opposing chirality, such that a generic solution can preserve ${\cal N}=(p,q)$ supersymmetry, meaning that it supports $2p$ real supercharges\footnote{$p$ Poincar\'e charges and $p$ conformal charges} of left chirality and a further $2q$ of right chirality. Second there is the compatibility of CFTs in two dimensions with the Virasoro algebra which allows for several distinct algebras realising a given number of chiral supercharges, at least when $p$ or $q$ become larger than $3$. These are classified in terms of the R-symmetry of the algebra G$_R$ and some representation $\rho$ of the lie algebra of G$_R$ that the currents transform in. 

The total number of supercharges an AdS$_3$ solution can support is constrained as  $p+q\leq 8$ \cite{Haupt:2018gap}, making a total of $16$ real supercharges maximal (half of what is possible in general within supergravity). The best understood realisations of the AdS/CFT correspondence are those which preserve maximal supersymmetry and admit an embedding into string theory. There is thus clear motivation to classify, and where possible, explicitly construct all maximally supersymmetric AdS$_3$ solution in ten and eleven dimensional supergravity. Efforts towards this goal include a classification of all solutions preserving  ${\cal N}=(4,4)$ supersymmetry in terms of the large superconformal algebra \cite{Sevrin:1988ew}  across \cite{DHoker:2008lup,DHoker:2008rje,DHoker:2009wlx,Estes:2012vm,Bachas:2013vza}\footnote{Note that the final paper \cite{Bachas:2013vza} generalises the earlier works in part to accommodate an interesting Janus solution dual to a $d=4$ interface found earlier in  \cite{Bobev:2013yra}} (dealing with $d=11$) and \cite{Macpherson:2018mif} (dealing with type II), while the local form of all solutions supporting the 4 distinct algebras  with ${\cal N}=(8,0)$ supersymmetry were constructed in \cite{Legramandi:2019xqd} (see also \cite{Dibitetto:2018ftj} for an earlier example and \cite{Deger:2019tem} for a gauged supergravity perspective). The status of generic ${\cal N}=(n,8-n)$ solutions is an open problem, though that of several cases can be inferred from existing chiral classifications\footnote{For instance the only solution preserving  ${\cal N}=(7,0)$ with algebra $\mathfrak{osp}(7|2)$ is locally AdS$_4\times \text{S}^7$ \cite{Macpherson:2023cbl}. Thus one knows that only one solution can possibly be compatible with  ${\cal N}=(7,1)$ with  algebra $\mathfrak{osp}(7|2)\oplus\mathfrak{osp}(1|2)$. In this case  AdS$_3$ gets enhanced to AdS$_4$ and so no true AdS$_3$ solution with these superconformal properties exists. }. 

A particularly interesting avenue to explore is the classification of solutions preserving ${\cal N}=(4,4)$ in terms of the small superconformal algebra. Several explicit examples are of course known, most famously the D1-D5 near horizon, in addition to  more recent constructions\cite{Dibitetto:2020bsh,Lozano:2022ouq}. While there has been some efforts towards the classification of solutions preserving small ${\cal N}=(4,0)$ supersymmetry \cite{Couzens:2017way,Lozano:2019emq,Macpherson:2022sbs}, the resulting classes are rather broad and yield little insight into when an enhancement to ${\cal N}=(4,4)$ happens. An interesting way to generate solutions with small ${\cal N}=(4,4)$ was presented in \cite{Capuozzo:2024onf} (see also \cite{Conti:2024rwd,Conti:2024qgx}), which realises an observation of \cite{Sevrin:1988ew} in supergravity. Starting from a solution with large $(4,4)$ one takes a slightly singular limit involving the associated continuous parameter in which the large algebra degenerates into the small one, however solutions generated in this fashion come with an additional flavour SO(4) not required for AdS$_3$ solutions with small $(4,4)$ in general, as such not all such solutions can be generated in this fashion\footnote{For example AdS$_3\times$S$^3\times K3$ certainly cannot.}. Thus the need for a classification focused on small (4,4) is clear and we initiate this process here by focusing on the case of $d=11$ supergravity.

A method that has proved extremely useful for constructing AdS$_3$ solutions with ${\cal N}=(n,0)$ supersymmetry has been to begin by constructing spinors that transform in the appropriate representation of the R-symmetry G$_R$ on a given background that supports bosonic fields that are G$_R$ singlets. When this is done it follows that whenever any ${\cal N}=(1,0)$ sub sector of these spinors solves the necessary conditions for supersymmetry, then they all do. This allows one to apply existing ${\cal N}=(1,0)$ G-structure conditions \cite{Dibitetto:2018ftj,Passias:2019rga,Passias:2020ubv,Macpherson:2021lbr} for the preservation of supersymmetry to the case of extended chiral symmetry. We would like to do something similar here,  but that requires G-structure conditions for ${\cal N}=(1,1)$ supersymmetric AdS$_3$, which are currently only known in type II supergravity \cite{Macpherson:2021lbr} - our first task then is to derive the $d=11$ analogue of these. Next it is necessary to construct general spinors that transform in the appropriate representation of the R-symmetry of the small superconformal algebra: We must construct two spinors on the $d=8$ internal space M$_8$ transforming in the \textbf{2}$\oplus \overline{\text{\textbf{2}}}$ of distinct SU(2) R-symmetries, spanning the $(4,0)$ and $(0,4)$ sectors of ${\cal N}=(4,4)$ - this necessitates that M$_8$ decomposes as a foliations of a round S$^3$ over some 5-manifold M$_5$. We can then use our $(1,1)$ G-structure conditions to classify the possible fluxes, warp factors and M$_5$'s - we will be able to give the local form of all such solutions up to solving Laplace equations. An important thing to appreciate is that the small ${\cal N}=(4,4)$ superconformal algebra is a sub algebra of its large analogue, where M$_5$ should further decompose in terms of a second 3-sphere and a Riemann surface - as such we are actually performing a classification of all ${\cal N}=(4,4)$ solutions in $d=11$. The local form of solutions preserving the large algebra was derived in \cite{Estes:2012vm,Bachas:2013vza}, we will also give an alternative form of these solutions expressed in terms of a single Laplace equation. \\
~\\
The lay out of the paper is as follows:

In section \ref{eq:neq11gstructures} we summarise the necessary and sufficient G-structure conditions  for ${\cal N}=(1,1)$ supersymmetric AdS$_3$ solutions in $d=11$ supergravity, which are derived in detail in appendices \ref{AdS$_3$ spinors and bi-linears} and \ref{d=11 N=(1,1)}. Similar to the AdS$_{2,3}$ classifications of \cite{Macpherson:2021lbr,Legramandi:2023fjr} we find that generically such solutions experience an enhancement to AdS$_4$. True AdS$_3$ solutions require some additional constraints to be imposed and, when they are, M$_8$ in general supports an SU(3)-structure. 

Section \ref{sec: classification} deals with the classification of (4,4) supersymmetric AdS$_3$. We construct spinors transforming in the appropriate representations of the R-symmetry and use them to construct the bi-linears that span a representative ${\cal N}=(1,1)$ SU(3)-structure on S$^3\times$M$_5$ in sub-section \ref{sec: gstructuresneq44}. We give a prescription for identifying when a class of solutions supports large rather than small superconformal symmetry, which should come with an additional 3-sphere, in sub-section  \ref{sec: vecorbilinears}. We then identify the independent classes of solution for which supersymmetry does and does not impose the presence of a second 3-sphere (whose presence does not guarantee large ${\cal N}=(4,4)$) in sub-sections \ref{sec:3-sphereorno3-sphere}. 

In section \ref{eq:smallclasses} we derive the local form (up to PDEs) of the 3 classes of local solutions that realise small (4,4) supersymmetry. For the first, presented in sub-section \ref{eq: R4class}, M$_5$ decomposes as warped $\mathbb{R}^4\times \mathbb{R}^1$ locally and solution are defined in terms of a degree  one polynomial $u$ on $\mathbb{R}^1$ and a generalised Laplace equation of M$_5$. For the second, presented in section \ref{eq: T3class}, M$_5$ decomposes in terms of a warped product of $\mathbb{T}^3$ and a Riemann surface with solutions defined in terms of an axially symmetric cylindrical Laplace equation. The third class of section \ref{eq:wickrotatedclass} turns out to be a subclass of the first, with 3-spherical symmetry in the $\mathbb{R}^4$ factor, and was previously found in \cite{Dibitetto:2020bsh}.

In section \ref{largeclass} we derive the 1 class of solutions that preserves  large ${\cal N}=(4,4)$ supersymmetry. In general such solutions are defined in terms of a parameter $\gamma$ and a Toda equation in two dimensions, however for all but one solution (namely AdS$_3\times$S$^3\times$S$^3\times \mathbb{T}^2$) this can be mapped to the same 3 dimensional cylindrical Laplace equation as section \ref{eq: T3class}. We establish that the classes of sections \ref{eq: T3class} and \ref{eq:wickrotatedclass}  (with $u=1$) can be extracted as certain limits of this class that involve $\gamma$ and also explain how our class relates to that of \cite{Bachas:2013vza}.

In section \ref{sec:examples} we find some new explicit examples of solutions that lie with the various classes we derive as well as recovering some existing solutions. Specifically we first construct a new family of solutions  with small ${\cal N}=(4,4)$ supersymmetry that are foliations of AdS$_3\times \text{S}^3\times \mathbb{T}^4$ over an interval. These are bounded between O2 planes and either M2 branes or a regular zero with an arbitrary number of smeared M5 branes placed along the interior. Next we  find a small (4,4) solution that fully back-reacts an M5 brane and O5 plane on AdS$_3\times$S$^3\times \mathbb{T}^5$. We then a construct a two parameter family of local solutions with large (4,4) supersymmetry from which we are able to recover 2 Asymptotically AdS$_7$ solutions recently found in \cite{Conti:2024rwd} for certain tunings of the parameters. Finally we construct a 3 parameter family of large (4,4) solutions which contains the asymptotically AdS$_4$ Janus solution of \cite{Bobev:2013yra}.

The main text closes with section \ref{sec: conclusions} where we present our conclusions after which there are several appendices which the main text refers to.

 \section{${\cal N}=(1,1)$ supersymmetric AdS$_3$ in $d=11$ supergravity}\label{eq:neq11gstructures}
In this section we present necessary and sufficient conditions for an AdS$_3$  solution in $d=11$ supergravity preserving ${\cal{N}}=(1,1)$ supersymmetry. The derivation of these results can be found in appendices \ref{AdS$_3$ spinors and bi-linears} and \ref{d=11 N=(1,1)}.\\
~\\
By definition an AdS$_3$ solution in $d=11$ supergravity (see appendix \ref{Conventions} for our conventions) must respect the isometries of AdS$_3$ and as such must have a metric and 4-form $G$ that are decomposable as
\begin{align}
\label{Ansatz}
 ds^2 = e^{2 A} ds^2 (\text{AdS}_3) + ds^2 (\text{M}_8), \qquad G = e^{3A} \text{vol} (\text{AdS}_3) \wedge G_1 + G_4.
\end{align}
Here we define the metric of AdS$_3$  such that its Ricci tensor is R$(\text{AdS$_3$}) = - 2 m^2 g(\text{AdS$_3$})$, where $m$ is the inverse radius of AdS$_3$. The AdS$_3$ warp factor $ e^{2A}$, the 1-form $G_1$ and the 4-form $G_4$ have support on M$_8$ only, such that the SO(2,2) isometry of AdS$_3$ is preserved.

The Bianchi identity and equations of motion of $G$ decompose into independent conditions orthogonal to and parallel to vol(AdS$_3$), respectively their magnetic portions are
\begin{align}
\label{magnetic}
d G_4 = 0, \qquad d\star_8 G_1 + \frac{1}{2} G_4 \wedge G_4 = 0,
\end{align}
while the electric parts are
\begin{align}
\label{electric}
d(e^{3A}G_1) = 0, \qquad d\left( e^{3A} \star_8 G_4 \right) - e^{3A} G_1 \wedge G_4 = 0,
\end{align}
where $\star_8$ is the Hodge dual on M$_8$.\\
~\\
In order to have a solution that preserves supersymmetry, \eqref{Ansatz} should support a $d=11$ Killing spinor $\epsilon$ that is Majorana and obeys the Killing spinor equation \eqref{eq:deq10KSE}. Given that our bosonic fields decompose as the (warped) product AdS$_3\times \text{M}_8$ we can take $\epsilon$ to decompose in terms of Killing spinors on AdS$_3$ $\zeta^{\pm}$ satisfying the following equation
\begin{align}
\label{KillingAdS}
\nabla_{\mu}\zeta^{\pm}=\pm \frac{m}{2}\gamma_{\mu} \zeta^{\pm}.
\end{align}
To realise ${\cal N}=(1,1)$ supersymmetry we must take the $d=11$ spinor to decompose as
\begin{align}
 \epsilon = \zeta^+ \otimes \chi^+ + \zeta^- \otimes \chi^-,\label{eq:neq11spinors}
\end{align}
where $ \left( \chi^+, \chi^- \right)$ are two non chiral Majorana spinors on M$_8$, which cannot vanish\footnote{strictly speaking one can vanish but then one would only have either (0,1) or (1,0) supersymmetry, which is not what we are interested in here.}.

From \eqref{eq:neq11spinors} and \eqref{KillingAdS} it is possible to construct G-structure conditions on M$_8$ that imply the preservation of ${\cal N}=(1,1)$ supersymmetry, as shown in appendix \ref{d=11 N=(1,1)}. There, instead of solving \eqref{eq:deq10KSE} directly, we make use of equivalent geometric condition for the preservation of supersymmetry on totally generic $d=11$ backgrounds derived in \cite{Gauntlett:2002fz}. These consist of differential constraints \eqref{fluxconstraints} on 3 distinct forms in ten dimensions whose components are defined by spinor bilinears as defined in \eqref{eq:10dforms}.  

An interesting outcome of our derivation, as shown between \eqref{firstdefinitionofK} and \eqref{eq:noads3ex}, is that ${\cal N}=(1,1)$ supersymmetric AdS$_3$ solutions necessarily experience an enhancement to AdS$_4$ unless one imposes 
\begin{align}
\label{constraintsinsection2}
& \xi =0, \qquad (\chi^{- \dagger} \chi^+ + \chi^{+ \dagger} \chi^-) = 0, \qquad ||\chi^{\pm}||^2 = e^A c,
\end{align}
where $\xi$ is a 1-form defined in \eqref{definitionxi} and $c$ is a constant - a similar result was previously found in \cite{Macpherson:2021lbr} for type II solutions with ${\cal N}=(1,1)$ supersymmetry. So to realise true AdS$_3$ solutions we must impose \eqref{constraintsinsection2}.

To solve \eqref{constraintsinsection2} we find it useful to decompose the internal metric as
\beq
ds^2(\text{M}_8)= ds^2(\text{M}_7)+ V^2,
\eeq
for $V$ a vielbein direction. We likewise decompose the spinors $(\chi^+, \chi^-)$ in terms of a  Majorana spinor $ \eta $ and unit norm 1-form $U$ on M$_7$  as
\begin{equation}\label{eq:deq7p1}
\chi^+ = \sqrt{c e^A}
\begin{pmatrix}
\sin \left( \frac{\alpha}{2} \right) \\
\cos \left( \frac{\alpha}{2} \right) \\
\end{pmatrix} \otimes \eta, \qquad
\chi^- =  - i \sqrt{c e^A}
\begin{pmatrix}
\sin \left( \frac{\alpha}{2} \right) \\
- \cos \left( \frac{\alpha}{2} \right) \\
\end{pmatrix} \otimes U \eta \qquad
U \cdot U = 1.
\end{equation}
We split the $d=8$ gamma matrices such that $\gamma_8$ corresponds to $V$ by taking $ \gamma_{a_7} = \sigma_2 \otimes \gamma^{(7)}_{a_7},\gamma_8 = \sigma_1 \otimes \mathbbm{1}$. The intertwiner is $B_8 = \mathbbm{1} \otimes B_7$ such that $ B_7^{-1} \gamma^{(7)}_a B_7 = - (\gamma^{(7)}_a)^{*}$ and we work in where convention $i \gamma^{(7)}_{1234567} = \mathbbm{1}$. \\
~\\
The spinor $\eta$, being Majorana and 7 dimensional defines a G$_2$-structure \cite{Gauntlett:2003cy}, with an associated real 3-form $\Phi_3$ which can be decomposed in terms of a unit norm 1-form $U$, a real 2-form $J$ and holomorphic 3-form $\Omega$ (both orthogonal to $U$) as
\begin{subequations}
\begin{align}
& \Phi_3 = - \frac{i}{3!} \eta^{\dagger} \gamma^{(7)}_{a_1 a_2 a_3} \eta e^{a_1 a_2 a_3} = J \wedge U - \text{Im} \Omega, \\[2mm]
& \star_7 \Phi_3 = \frac{1}{2} J \wedge J + \text{Re} \Omega \wedge U,
\end{align}
\end{subequations}
where $e^{a}$ is a vielbein basis on  M$_7$ and $(U,J,\Omega)$ are orthogonal to $V$.  Together $(J,\Omega)$ span an SU(3)-structure on the 6-dimensional space orthogonal to  $(U,V)$ which one can take to be a vielbein on the remaining 2 dimensions in M$_8$. In general SU(3)-structure forms in $d=6$ must obey
\beq
J\wedge J\wedge J=\frac{3i}{4} \Omega \wedge\overline{\Omega}=6\text{vol}_6,~~~J\wedge \Omega=0,
\eeq
but other than being orthogonal to $(U,V)$ are otherwise unconstrained. 

As shown in appendix \ref{d=11 N=(1,1)} the spinor bilinears one can construct from $(\chi^+,\chi^-)$, give rise to form on M$_8$, defined in \eqref{definitionsofSU(3)forms}, which are spanned by $(\alpha, U,V,J,\Omega)$. The ten dimensional forms of \cite{Gauntlett:2002fz} then decompose in terms of wedge products of these and additional forms on AdS$_3$ defined in appendix \ref{AdS$_3$ spinors and bi-linears}. After plugging these into the $d=10$ geometric conditions for supersymmetry, one can factor out the AdS$_3$ dependence and arrive at system of conditions on M$_8$ that imply ${\cal N}=(1,1)$  supersymmetry for AdS$_3$. These can most succinctly be expressed in terms of the complex 1-form
\begin{align}
Z = U + i V.
\end{align} 

In summary necessary and sufficient conditions for ${\cal{N}}=(1,1)$ supersymmetry are given by 
\begin{subequations}
\begin{align}
& d\left( e^{2A} \sin \alpha Z \right) - m e^A \left( J + \cos \alpha U \wedge V \right) = 0,\label{BPS-conditionscomplex1} \\[2mm]
& e^{3A} G_1 + 2 e^{2A} m \sin \alpha V - d \left(e^{3A} \cos \alpha \right) = 0,\label{BPS-conditionscomplex2} \\[2mm]
& d(e^A \sin \alpha \text{Im} \Omega \wedge U \wedge V) + e^A (J + \cos \alpha U \wedge V)\wedge G_4 = 0,\label{BPS-conditionscomplex3} \\[2mm]
& d\left(e^{2A} (\cos \alpha \text{Im} \Omega - i \text{Re} \Omega )\wedge Z \right)- e^{2A} \sin \alpha Z \wedge G_4 - e^A m \sin \alpha \text{Im} \Omega \wedge U \wedge V = 0, \label{BPS-conditionscomplex4}\\[2mm]
& d\left(e^{3A} \sin \alpha \text{Im} \Omega \right) - e^{3A} (\star_8 G_4 - \cos \alpha G_4 )- 2 e^{2A} m (\cos \alpha \text{Im} \Omega \wedge V - \text{Re} \Omega \wedge U) = 0,\label{BPS-conditionscomplex5} \\[2mm]
& 6 \star_8 dA + \sin \alpha \text{Im} \Omega \wedge G_4 - 2 \cos \alpha \star_8 G_1 = 0,\label{BPS-conditionscomplex6} \\[2mm]
& e^A Z \wedge (2 \star_8 G_1 \sin \alpha + ( \text{Im} \Omega \cos \alpha - i \text{Re} \Omega ) \wedge G_4) -  6 m i \text{vol(M$_8$}) = 0.\label{BPS-conditionscomplex7}
\end{align}
\end{subequations}
Note that it follows from \eqref{BPS-conditionscomplex1} that we cannot fix $\sin\alpha=0$ without also fixing $m=0$, a  limit in which AdS$_3$ blows up into Minkowski$_3$.

While  \eqref{BPS-conditionscomplex1}-\eqref{BPS-conditionscomplex7} guarantee supersymmetry, they do not guarantee that one has a solution of  $d=11$ supergravity, for that by definition one must solve Einstein's equations and the Bianchi identity and equation of motion of the 4-form flux (ie in this case \eqref{magnetic} and \eqref{electric}). One can show that \eqref{BPS-conditionscomplex1}-\eqref{BPS-conditionscomplex7} imply \eqref{electric} when \eqref{magnetic} are assumed to hold. An integrability proof of \cite{Gauntlett:2002fz} then implies that one has an ${\cal N}=(1,1)$ supersymmetric AdS$_3$ solution when \eqref{BPS-conditionscomplex1}-\eqref{BPS-conditionscomplex7} are solved and \eqref{magnetic} hold\footnote{This follows because, as shown in appendix \ref{d=11 N=(1,1)}, the canonical Killing vector of \cite{Gauntlett:2002fz} is necessarily time-like for AdS$_2$ solutions for which AdS$_3$ is not merely embedded into AdS$_4$}.

\section{Classifying ${\cal N}=(4,4)$ AdS$_3$ solutions}\label{sec: classification}
Having derived necessary conditions for an AdS$_3$ solution of $d=11$ supergravity to preserve ${\cal N}=(1,1)$ supersymmetry in the previous section, in this section we shall put these to work by classifying solutions preserving  ${\cal N}=(4,4)$.
\subsection{G-structures for ${\cal N}=(4,4)$ AdS$_3$ solutions}\label{sec: gstructuresneq44}
An AdS$_3$ solution that preserves ${\cal N}=(4,4)$ must support both a left and right chiral superconformal algebra $\mathfrak{G}_{\pm}$ each of which are characterised by an R-symmetry group G$^{\pm}_{R}$ and  representation $\rho^{\pm}$ in which the fermionic generators transform. There are two distinct  ${\cal N}=4$ chiral algebras a solution can realise
\begin{itemize}
\item $\mathfrak{G}=\mathfrak{su}(1,1|2)$/$\mathfrak{u}(1)$: Small ${\cal N}=4$ with G$_R=$ SU(2) and $\rho=$ $\text{\textbf{2}}\oplus \overline{\text{\textbf{2}}}$.\\
\item $\mathfrak{G}=\mathfrak{d}(2,1;\alpha)$: Large ${\cal N}=4$ with G$_R=$ SO(4) and $\rho=$ \textbf{(2,2)},
\end{itemize}
where $\mathfrak{su}(1,1|2)$/$\mathfrak{u}(1)$ has a single SU(2) current algebra of level $k$ associated to it while $\mathfrak{d}(2,1;\alpha)$ has two SU(2) current algebras of levels $k_{\pm}$ with respect to which the  parameter $\alpha=\frac{k_+}{k_-}$. The large algebra actually contains several others for certain tunings of $\alpha$, ie $\mathfrak{d}(2,1;1)=\mathfrak{osp}(4|2)$. It is also possible to realise the small superconformal as a slightly singular limit of the large one, however it is not the case that all solutions supporting the former algebra can be obtained as limits of solutions supporting the latter.

One can arrange for an AdS$_3$ solution to preserve a given algebra by ensuring that its bosonic fields are singlets with respect to the required R-symmetry group and that it supports Killing spinors that transform in the corresponding representation. In the case at hand we should take
\begin{align}
 \epsilon = \sum_{I=1}^4\zeta^+ \otimes \chi^{I+} + \zeta^- \otimes \chi^{I-},\label{eq:neq44spinors}
\end{align}
such that $\chi^{I\pm}$ transform in the $\rho^{\pm}$ representation of $G^{\pm}_R$ whilst simultaneously being singlets with respect to $G^{\mp}_R$. This requirement is restrictive, and while there are many ways to realise a single chiral ${\cal N}=4$ superconformal algebra, it appears that there are only two options for realising ${\cal N}=(4,4)$, with two copies of the large or two copies of the small ${\cal N}=4$ superconformal algebras\footnote{The reason one does to expect to see, say, a solution preserving small $(4,0)$ and large $(0,4)$ simultaneously is the need for the left/right spinors to be singlets with respect to the right/left R-symmetry. That means you need an internal space that supports two sets of spinors, one in the (\textbf{2} $\oplus$ $\overline{\text{\textbf{2}}}$,\textbf{1},\textbf{1}) and one in the (\textbf{1},\textbf{2},\textbf{2}) of SU(2)$\times$SU(2)$\times$SU(2). The smallest spaces realising this are $\text{S}^3\times\text{S}^3\times\text{S}^3$ and SU(2)$^3$ preserving squashings and/or fiberings there of, these are too large to embed into $d=11$ supergravity with an AdS$_3$ factor.}. To realise the large ${\cal N}=(4,4)$ case it is necessary for the solution to contain two 3-spheres to realise the requisite  SO(4)$\times$SO(4) R-symmetry, though this is not sufficient, and it is possible for such solutions to only preserve small ${\cal N}=(4,4)$. To realise the small ${\cal N}=(4,4)$ case only a single 3-sphere is required, making solutions supporting the large algebra a subset of those supporting the small one. We thus see that to realise ${\cal N}=(4,4)$ supersymmetry we must refine the bosonic fields of \eqref{Ansatz} as 
\beq
ds^2(\text{M}_8)= e^{2C}ds^2(\text{S}^3)+ds^2(\text{M}_5),~~~~G_4= e^{3C} \text{vol}(\text{S}^3)\wedge g_1+g_4,
\eeq
where $(e^A,e^C,G_1,g_1,g_4)$ have support on M$_5$ only. As explored in appendix \ref{sec:S3} the 3-sphere supports two sets of Killing spinors $\xi^{\pm}$  which obey
\beq
\nabla_{a_3}^{\text{S}^3}\xi^{\pm}=\pm \frac{i}{2}\gamma_{a_3}\xi^{\pm}.
\eeq
The 3-sphere has an SO(4) $\sim$ SU(2)$_+\times$SU(2)$_-$ global symmetry with respect to which $\xi^{\pm}$ transforms in the \textbf{2} of SU(2)$_{\pm}$ and \textbf{1} of SU(2)$_{\mp}$, we shall thus decompose $\chi^{I\pm}$ in terms of $\xi^{\pm}$.  

Each of the 4 independent ${\cal N}=(1,1)$ Killing spinors contained in \eqref{eq:neq44spinors} must solve an independent version of \eqref{BPS-conditionscomplex1}-\eqref{BPS-conditionscomplex7} for the same $(e^A,e^C,G_1,g_1,g_4)$, and each of these has a closed 1-form  $e^{2A}\sin\alpha V$ associated to it.  As the 1-form bi-linears formed from  $(\xi^{\pm},\xi^{\pm})$ are parallel to  the SU(2) invariant forms (see appendix \ref{sec:S3}), and $e^{2A}\sin\alpha V$, is defined in terms of $(\chi^{I\pm},\chi^{I\pm})$ we must impose that the S$^3$ data lies orthogonal to this 1-form. In particular this means we can use the $d=7+1$ decomposition of the spinor in \eqref{eq:deq7p1} assuming that $\xi^{\pm}$ lies entirely within the $d=7$ spinors without loss of generality.

We find it useful to further decompose the $d=7$ gamma matrices in terms of gamma matrices on the 3-sphere and the space orthogonal to  as
\beq
\gamma^{(7)}_{a_3}=\sigma_{a_3}\otimes \hat\gamma^{(4)},~~~~\gamma^{(7)}_{a_4}=\mathbbm{1}\otimes \gamma^{(4)}_{a_4},~~~~B_7=\sigma_2\otimes B_4,
\eeq
where $\gamma^{(4)}_{a_{4}}$ are gamma matrices in 4 dimensions, $\sigma_{a_3}$ are the Pauli matrices,  $\hat\gamma^{(4)}=-\gamma^{(4)}_{1234}$ is the $d=4$ chirality matrix  and $B_4$ is such that $B_4 B_4^{*}=- B_4 B_4=1$. Following the recipe in  \cite{Lozano:2019emq}, general $d=8$ Majorana spinors that transform in the \textbf{2}$\oplus\overline{\text{\textbf{2}}}$ of SU(2)$_{\pm}$ are then given by
\begin{equation}
\begin{split}
\chi^{I+} &= \sqrt{c e^A}
\left(\begin{array}{c}
\sin \left( \frac{\alpha}{2} \right) \\
\cos \left( \frac{\alpha}{2} \right) \\
\end{array}\right) \otimes \bigg[Y^I_{\mathfrak{ab}}\xi^{\mathfrak{a}+}\otimes\eta^{\mathfrak{b}+}\bigg], \\[2mm]
\chi^{I-} &=  \sqrt{c e^A}
\left(\begin{array}{c}
\sin \left( \frac{\alpha}{2} \right) \\
- \cos \left( \frac{\alpha}{2} \right) \\
\end{array}\right) \otimes\bigg[ Y^I_{\mathfrak{ab}}\xi^{\mathfrak{a}+}\otimes\eta^{\mathfrak{b}-}\bigg], \\[2mm]
Y^I&=(\sigma_2\sigma_1,\sigma_2 \sigma_2,\sigma_2\sigma_3,-i \sigma_2),\label{eq:deq7p122}
\end{split}
\end{equation}
where we define the following doublets of spinors on S$^3$ and  the remaining $d=4$ space
\begin{equation}
\xi^{\mathfrak{a}\pm} = \begin{pmatrix}
\xi^{\pm} \\
\xi^{\pm c} \\
\end{pmatrix}^{\mathfrak{a}},~~~~ \eta^{\mathfrak{a}\pm} = \begin{pmatrix}
\eta^{\pm} \\
\eta^{\pm c} \\
\end{pmatrix}^{\mathfrak{a}},
\end{equation}
where the superscript $c$ stands for Majorana conjugate and we take $\xi^{\pm}$ to have unit norm.  The two $d=4$ spinors can in general be expanded in a common basis in terms of two unit norm chiral spinors $\eta_{\pm}$ as
\beq
\eta^+=\cos\left(\frac{\beta}{2}\right) \eta_++\sin\left(\frac{\beta}{2}\right) \eta_-,~~~~\eta^-= a_1 \eta_+ + a_2 \eta^c_+ + a_3 \eta_- + a_4 \eta^c_-,~~~~|a_1|^2+|a_2|^2+|a_3|^2+|a_4|^2=1,
\eeq
where $(\beta, a_1 , a_2 , a_2 , a_3 )$ are functions with support on M$_4$. The chiral spinors give rise to a vielbein on M$_4$ as
\begin{equation}
\begin{split}
\eta^{\dag}_{\mp}\gamma_a^{(4)}\eta_{\pm}&= w^1\pm i w^2,~~~~\eta^{c\dag}_{\mp}\gamma_a^{(4)}\eta_{\pm}=-(w^3+i w^4), \\[2mm]
ds^2(\text{M}_5)&=ds^2(\text{M}_4)+V^2,~~~~~ds^2(\text{M}_4)=(w^1)^2+(w^2)^2+(w^3)^2+(w^4)^3.
\end{split}
\end{equation}
 We still need to solve \eqref{constraintsinsection2} which fixes the functions of the spinor ansatz in terms of $\beta$ and 4 real functions $\rho_{1,2,3,4}$ constrained such that $(\rho_1)^2+(\rho_2)^2+(\rho_3)^2+(\rho_4)^2=1$ as
\begin{equation}
\begin{split}
a_1&=\sin\left(\frac{\beta}{2}\right)(\rho_1+i\rho_4),~~~~a_2=\sin\left(\frac{\beta}{2}\right)(\rho_3+i\rho_2), \\[2mm]
a_3&=-\cos\left(\frac{\beta}{2}\right)(\rho_1+i\rho_4),~~~a_4= -\cos\left(\frac{\beta}{2}\right)(\rho_3+i\rho_2).
\end{split}
\end{equation}
We now have two sets of spinors $\chi^{I\pm}$ in $d=8$ that transform in \textbf{2}$\oplus \overline{\text{\textbf{2}}}$ of SU(2)$_{\pm}$ and solve the conditions which ensure that the solutions they support are not locally AdS$_4$, we now want to use the ${\cal N}=(1,1)$ G-structure conditions of the previous section to find the geometries consistent with them.\\
~\\
Because \eqref{eq:deq7p122} decomposes in terms of 4 independent ${\cal N}=(1,1)$ subsectors, if the SU(3)-structure forms that follow from each of these solve \eqref{BPS-conditionscomplex1}-\eqref{BPS-conditionscomplex7} for the same bosonic fields\\ $(e^A,e^C,ds^2(\text{M}_5), G_1,g_1,g_4 )$ then we preserve ${\cal N}=(4,4)$ supersymmetry.  Manifestly the  bosonic fields of our ansatz are SU(2)$_+\times$SU(2)$_-$ singlets, while the spinors $\zeta^{\pm}\otimes \chi^{I \pm}$ transform in the  \textbf{2}$\oplus \overline{\text{\textbf{2}}}$ of SU(2)$_{\pm}$. This fact ensures that if we solve \eqref{BPS-conditionscomplex1}-\eqref{BPS-conditionscomplex7} for a single ${\cal N}=(1,1)$ subsector then the other 3 are implied through the action of SU(2)$_{\pm}$. This is perhaps easier to see in terms of the Killing spinor equations the background must satisfy, a proof is given in section 2 of\cite{Legramandi:2020txf}. Going forward then, we shall choose our ${\cal N}=(1,1)$ subsector to be
\beq
\chi^{+}=\chi^{4+},~~~~\chi^{-}=\chi^{4-}.
\eeq
Through the formulae in \eqref{definitionsofSU(3)forms} it is then a simple matter to compute the SU(3)-structure forms which appear in the necessary conditions for ${\cal N}=(1,1)$ supersymmetry in \eqref{BPS-conditionscomplex1}-\eqref{BPS-conditionscomplex7}. \\
~\\
To present the SU(3)-structure forms we find it useful to decompose the 3-sphere metric as
\beq
ds^2(\text{S}^3)= d\theta^2+ \sin^2\theta ds^2(\text{S}^2),~~~~\text{vol}(\text{S}^3)=\sin^2\theta d\theta\wedge \text{vol}(\text{S}^2).
\eeq
This  2-sphere  supports an SU(2)$_D$ isometry which is the anti-diagonal subgroup of SU(2)$_+\times$SU(2)$_-$. Introducing a set of unit norm embedding coordinates for the 2-sphere,  $y_i$  such that $ds^2(\text{S}^2)=(dy_i)^2$, we have the following SU(2)$_D$ triplets
\beq
y_i,~~~dy_i,~~~~y_i\text{vol}(\text{S}^2),~~~K_{i}=\epsilon_{ijk}y_i dy_j,
\eeq
which form a closed set under $d$ and $\wedge$ through the identities
\beq
dK_i=2y_i\text{vol}(\text{S}^2),~~~~dy_i\wedge K_i=2\text{vol}(\text{S}^2),~~~~\epsilon_{ijk}K_j\wedge dy_k=0.
\eeq
We also find it useful to introduce the following combinations of one forms on M$_4$ as
\beq
\mu^i=(w^3,-w^4,-w^1),~~~E^i=\rho_1 \mu^i-\epsilon_{ijk} \rho_j \mu^k,~~~\hat E^i= \mu^i-2 \epsilon_{ijk}\rho_j E^k,~~~W=w^2.\label{eq: tripledefs}
\eeq
we also define $\rho_i=(\rho_2,\rho_3,\rho_4)_i$.
In terms of the above the SU(3)-structure forms defined on M$_7$ then decompose as
\begin{align}
U&=  - e^C \rho_1 \sin \beta d(\cos \theta )-e^C \sin (\beta ) \rho_i d(y_i \sin \theta )-\cos \theta (\cos \beta \mu^i \rho_i + \rho_1 W) \nn\\ 
& +\cos \beta (y_i \sin \theta)E^i-W \rho_i (y_i \sin \theta), \nn \\
J &= - d(\cos \theta) \wedge (e^C (\rho_i \mu^i + \rho_1 W \cos\beta )) - \cos\theta W\wedge \sin\beta \rho_i \mu^i \nn \\
& - d(\sin \theta y_i )\wedge ( e^C (W \cos \beta  \rho_i - E^i))-\sin (\theta ) W\wedge (\sin \beta E^i y_i), \nn\\[2mm]
\text{Re} \Omega &= -\cos \beta   (\rho_1 \cos \theta +\sin \theta \rho_i y_i) \mu^1 \wedge \mu^2 \wedge \mu^3+ e^{2 C} \sin^3\theta \text{vol}(\text{S}^2) \wedge ( \rho_1 W + \cos \beta \rho_i \mu^i ) \notag \nn\\
&  +  e^{3 C} \sin \beta \left(\rho_1  \cos \theta + \sin \theta y_i \rho_i \right) \text{vol}(\text{S}^3)+ \frac{1}{2} \epsilon_{ijk} \biggl( d(\cos \theta )\wedge (e^C \sin \beta \rho_i \mu^j \wedge \mu^k) \notag \\
& + \cos \theta  \rho_i W \wedge \mu^j \wedge \mu^k + d(\sin \theta  y_i )\wedge (e^C \sin \beta (\rho_1 \mu^j-2 E^j )\wedge \mu^k ) \nn\\ 
& \sin \theta y_i W \wedge (\rho_1 \mu^j - 2 E^j )\wedge \mu^k\biggr) +e^{2C} \left(\text{vol}(\text{S}^2) \sin ^2\theta \cos \theta y_i - d(\cos \theta) \wedge K_{_i} \right) \wedge (\cos \beta E^i - W \rho_i ), \notag \\[2mm]
\text{Im} \Omega & = - e^{3 C} \cos \beta \text{vol}(\text{S}^3) - \sin \beta \mu^1 \wedge \mu^2 \wedge \mu^3 + \frac{1}{2} \epsilon_{ijk} \biggl( d\theta \wedge (y_i e^C \cos \beta E^j \wedge E^k )\nn \\
& +\sin \theta \cos \theta dy_i \wedge (e^C \cos \beta E^j \wedge E^k )  + \sin ^2\theta K_{i} \wedge (e^C \cos \beta (\mu^j \wedge \mu^k -E^j \wedge E^k)) \biggr) \nn \\
& + \frac{1}{2} \biggl( \sin ^2\theta d\theta \wedge dy_i \wedge \left(e^{2C} \sin \beta (\hat{E}^i-\mu^i)\right) + \cos \theta \sin \theta   d\theta \wedge K_{i}  \wedge \left(e^{2 C} \sin \beta (\mu^i + \hat{E}^i)\right) \notag \nn\\
& + \sin ^2\theta y_i \text{vol}(\text{S}^2) \wedge \left(e^{2C} \sin \beta (\hat{E}^i +\mu^i ) \right) + \cos \theta \sin \theta dy_i \wedge (e^C W \wedge (\hat{E}^i-\mu^i )) \nn \\
& - \sin^2\theta K_i \wedge (e^C W \wedge (\mu^i + \hat{E}^i)) - y_i d\theta \wedge (e^C W \wedge (\mu^i - \hat{E}^i))\biggr). \label{eq:SU3forms} 
\end{align}
We have already argued that the above SU(3)-structure forms should give rise  to  solutions preserving both the large and small ${\cal N}=(4,4)$ superconformal algebras when they solve \eqref{BPS-conditionscomplex1}-\eqref{BPS-conditionscomplex7}. However there is a way to get some insight into when to expect which algebra without first solving \eqref{BPS-conditionscomplex1}-\eqref{BPS-conditionscomplex7}, as we explore in the next section.
\subsection{General Killing vector bi-linears and which algebra to expect} \label{sec: vecorbilinears}
In \cite{Couzens:2022agr} it was shown that certain vector bi-linears one can define on the internal space of AdS$_3$ solutions preserving chiral extended supersymmetry are necessarily Killing vectors. Lifting this result to $d=11$ supergravity we find that the following 1-form bilinears must be dual to Killing vectors on M$_8$ in general
\beq
{\cal K}^{ IJ\pm}=\chi^{I\pm}\gamma^{(8)}_a\hat\gamma\chi^{J\pm}e^a.\label{eq:killingmatrixgen}
\eeq
For our specific spinor ansatz and decomposition of M$_8$ these become\footnote{Here we decided to leave implicit the indices $IJ$ of the one-form ${\cal K}^{IJ\pm}$. It is understood these  are given by structure of the matrices $\tensor{\Sigma}{^{i}^{IJ}}$, $\tensor{\tilde{\Sigma}}{^{i}^{IJ}}$. From here on out, we  write only ${\cal K}^{\pm}$, $\tensor{\Sigma}{^{i}}$, $\tensor{\tilde{\Sigma}}{^{i}}$ hiding $IJ$ for the sake of convention.}
\begin{align}
{\cal K}^{+}&= e^{A}\sin\alpha c\left(e^{C}\cos\beta k^{+}_i\Sigma^i-2\sin\beta \mu^i\tilde{\Sigma}^i\right),\nn\\[2mm]
{\cal K}^{-}&= e^{A}\sin\alpha c\left(e^{C}\cos\beta k^{-}_i\Sigma^i-2\sin\beta \hat E^i\tilde{\Sigma}^i\right).\label{eq:Killingmatrices}
\end{align}
where $k^{\pm}_i$ are a triplet of SU(2) forms on the 3-sphere charged under SU(2)$_{\pm}$ as defined in appendix \ref{sec:S3} and $(\Sigma^{i},\tilde{\Sigma}^i)$ are two mutually commuting real representations of the Lie algebra of SU(2) in 4 dimensions, specifically
\beq
\Sigma^i= \frac{1}{2}(i\sigma_2\otimes \sigma_1,-i\sigma_2\otimes\sigma_3,i\mathbbm{1}\otimes \sigma_2)^i,~~~~\tilde\Sigma^i= \frac{1}{2}(i\sigma_1\otimes \sigma_2,i\sigma_2\otimes \mathbbm{1},-i\sigma_3\otimes \sigma_2)^i.
\eeq
That \eqref{eq:Killingmatrices} must be dual to Killing vectors informs us that, if they are non-zero, \\$(e^{A}\sin\alpha \sin\beta \mu^i,~e^{A}\sin\alpha \sin\beta\hat E^i)$ are dual to Killing vectors on M$_4$ and that, if it is non-zero,  $e^{A+C}\sin\alpha\cos\beta \propto e^{2C}$. The reason this provides a hint as to what algebra is preserved is that, as argued in \cite{Macpherson:2022sbs}, all examples of AdS$_3$ solution that the type II analogue of \eqref{eq:killingmatrixgen} has been tested on are such that $K^{IJ \pm}$ contain all of the Killing vectors of the R-symmetry, although possibly also contain additional Killing vectors under which the spinors are singlets. This suggests that the large ${\cal N}=(4,4)$ superconformal algebra is only realised when $(\sin\beta,\cos\beta)\neq 0$ and $(\mu^i, \hat E^i)$ are both non-zero and propositional to independent SU(2) 1-forms $\tilde{k}^{\pm}_i$ on a second 3-sphere $\tilde{\text{S}}^3$ with all other situations giving rise to small ${\cal N}=(4,4)$ - this indeed turns out to be the case.

\subsection{Classes with and without an additional 3-sphere}\label{sec:3-sphereorno3-sphere}
The purpose of this section is to derive the system of conditions on M$_5$ that substituting \eqref{eq:SU3forms} into \eqref{BPS-conditionscomplex1}-\eqref{BPS-conditionscomplex7} leads to, ie the conditions on M$_5$ one gets after factoring out the 3-sphere data.\\ 
~~\\
Given that we seek solutions that preserve ${\cal N}=(4,4)$, a maximal case for solutions with an AdS$_3$ factor \cite{Haupt:2018gap} we expect the $d=8$ G-structure conditions \eqref{BPS-conditionscomplex1}-\eqref{BPS-conditionscomplex7} to contain many redundancies, indeed it turns out that \eqref{BPS-conditionscomplex6}-\eqref{BPS-conditionscomplex7} are implied by the rest for the ansatz we consider, so our focus here will be on \eqref{BPS-conditionscomplex1}-\eqref{BPS-conditionscomplex5}. Upon inserting the bi-linears of \eqref{eq:SU3forms} into the real part of \eqref{BPS-conditionscomplex1} we find two very important conditions that appear wedged with respectively $d(\cos\theta)$ and $d(\sin\theta y_i)$, namely
\begin{equation}
\begin{split}
P_1&=d\left(e^{2A+C}\sin\alpha \sin\beta \rho_1\right)-e^{2A}\sin\alpha(\rho_1 W+\cos\beta \rho_i \mu^i) \\[2mm]
&+e^{A+C}m\left(\rho_i\mu^i +\cos\beta\rho_1  W+ \rho_1 \cos\alpha \sin\beta V\right)=0, \\[2mm]
P_i&=d\left(e^{2A+C}\sin\alpha \sin\beta \rho_i\right)+e^{2A}\sin\alpha (-\rho_i W+\cos\beta E^i) \\[2mm]
&- e^{A+C}m\left(E^i-\cos\beta \rho_i W- \cos\alpha\sin\beta \rho_i V\right)=0.
\end{split}
\end{equation}
By taking the combination $\rho_1P_1+\rho_i P_i$ and given that $\rho_1^2+...+\rho_4^2=1$ we find
\beq
d(e^{2A+C}\sin\beta \sin\alpha)+e^{A}W(me^{C}\cos\beta-e^{A}\sin\alpha)+e^{A+C}m\cos\alpha\sin\beta V\label{eq:intercond}.
\eeq
If we then combine $P_i-\rho_i$\eqref{eq:intercond} we arrive at
\beq
(e^C m-e^A \cos\beta \sin\alpha)E^i= e^{A+C}\sin\alpha \sin\beta d\rho_i,
\eeq
where we remind the reader that $\sin\alpha=0$ is incompatible with AdS$_3$ as explained below \eqref{BPS-conditionscomplex7}.
As all components of $E^i$ are necessarily non-zero this condition gives a branching of possible solutions, ie either  $\sin\beta d\rho_i=0$ or $(e^C m-e^A \cos\beta \sin\alpha)\neq 0$ in which case we get a definition of 3 vielbein directions in M$_4$. At first sight it appears that $\sin\beta d\rho_i=0$ can be solved in 2 ways, however if one sets $\sin\beta=0$ in \eqref{eq:SU3forms}  one can then rotate out all the dependence on $(\rho_1,\rho_i)$ via an SO(4) rotation of $w^{1},...,w^4$ making $d\rho_i=0$ the more general choice than $\sin\beta=0$. Further if $d\rho_i=0$ it is possible to fix $(\rho_1,\rho_i)=(1,0)$ without loss of generality. To see this one one should note that \eqref{eq:SU3forms} are actually spanned by embedding coordinates on a round 3-sphere defined as
\beq
Y_A= (\cos\theta, \sin\theta y_i)_A.
\eeq
Then since $(\rho_1,\rho_i)$ are constant, we can effectively absorb them into the definition of $Y_A$ with an SO(4) rotation, making the precise values of $(\rho_1,\rho_i)$ irrelevant. Conversely, if we assume that  $\sin\beta d\rho_i \neq 0$ then we find that $E^i$ define a vielbein on a second 3-sphere $\tilde{\text{S}}^3$, ie
\beq
E^i= \frac{e^{A+C}\sin\alpha \sin\beta}{e^C m-e^A \cos\beta \sin\alpha} d\rho_i~~~~\Rightarrow~~~~  \mu^i \mu^i= \frac{e^{2(A+C)}\sin^2\alpha\sin^2\beta}{(e^C m-e^A \cos\beta \sin\alpha)^2}ds^2(\tilde{\text{S}}^3),
\eeq
where $ds^2(\tilde{\text{S}}^3)=d\rho_1^2+...+d\rho_4^2$, i.e in this case $(\rho_1,\rho_i)$ are embedding coordinates for a second 3-sphere. Given the above definition of $E^i$ one can the easily extract $(\mu^i,\hat E^i)$ from \eqref{eq: tripledefs}, we find
\beq
\mu^i= \frac{1}{2}\frac{e^{A+C}\sin\alpha\sin\beta}{e^C m-e^A \cos\beta \sin\alpha}\tilde{k}^+_i,~~~~\hat E^i= \frac{1}{2}\frac{e^{A+C}\sin\alpha\sin\beta}{e^C m-e^A \cos\beta \sin\alpha}\tilde{k}^-_i,
\eeq
where the specific form of $\tilde k^{\pm}_i$ can be extracted easily from \eqref{eq: tripledefs}, i.e. using that $\rho_1^2+(\rho_i)^2=1$ we can write them as
\beq
\tilde k^{\pm}_i=\left(\begin{array}{c}\rho_1d\rho_2-\rho_2d\rho_1\pm (\rho_3 d\rho_4-\rho_4 d\rho_3)\\
\rho_1 d\rho_3-\rho_3 d\rho_1\pm(\rho_4 d\rho_2-\rho_2 d\rho_4)\\
\rho_1 d\rho_4-\rho_4 d\rho_1\pm(\rho_2 d\rho_3-\rho_3 d\rho_2)\end{array}\right)
\eeq
but the important point is that they are such that $d\tilde{k}^{\pm}_i=\pm \frac{1}{2}\epsilon_{ijk}\tilde{k}^{\pm}_j\wedge\tilde{k}^{\pm}_k$, ie they are a second set of SU(2) triplets charged under the $\widetilde{\text{SU}}(2)_{\pm}$ subgroups of $\tilde{\text{S}}^3$.

We thus see that there are two classes of solution distinguished by the necessary existence, or not, of an additional 3-sphere $\tilde{\text{S}}^3$. 

\subsubsection{Case with no necessary extra 3-sphere}
As argued in the proceeding section, classes of solution with M$_5$ not necessarily containing an additional 3-sphere are such that one can fix
\beq
\rho_1=1,~~~~\rho_i=0,
\eeq 
without loss of generality. After factoring out the 3-sphere data from \eqref{BPS-conditionscomplex1}-\eqref{BPS-conditionscomplex5} one is left with a highly redundant system of conditions on M$_5$. Through significant work eliminating redundant conditions it is possible to show that this system is implied in general by
\begin{subequations}
\begin{align}
&m e^{C}-e^{A}\cos\beta \sin\alpha=0\label{eq:nos3bps1},\\[2mm]
& d(e^{2A+C}\sin\alpha \sin\beta)+ e^{2A}\sin\alpha \sin\beta(\cos\alpha \cos\beta V-\sin\beta W),\label{eq:nos3bps2}\\[2mm]
&d(e^{2A}\sin\alpha\sin\beta (\cos\alpha \cos\beta W+ \sin\beta V))=0,\label{eq:nos3bps2b}\\[2mm]
&d(e^{2A}\sin\alpha V)=0,\label{eq:nos3bps3}\\[2mm]
&d(e^{2A}\sin\alpha W)- me^{A} \cos\alpha W\wedge V=0,\label{eq:nos3bps4}\\[2mm]
&d(e^{A+C} \mu^i)+ e^{-C}(\cos\alpha\cos\beta V-\sin\beta W)\wedge (e^{A+C} \mu^i)\label{eq:nos3bps5},\\[2mm]
&d(\cos\beta )\wedge W\wedge V=0,\label{eq:nos3bps6}\\[2mm]
&d(e^{-C}(\cos\alpha \cos\beta W+ \sin\beta V))+e^{-2C}(\cos\alpha \cos\beta W+ \sin\beta V)\wedge (\cos\alpha \cos\beta V-\sin\beta W)=0,\label{eq:nos3bps7}\\[2mm]
& e^{3A} G_1 + 2 e^{2A} m \sin \alpha V - d \left(e^{3A} \cos \alpha \right) = 0,\label{eq:nos3bps8}\\[2mm]
&d(e^{A+2C}\cos\alpha)-\frac{2}{m} e^{2A}\sin\alpha \sin\beta(\cos\alpha \cos\beta W+\sin\beta V)+ \frac{2}{m}e^{2A}\sin\alpha V= m e^{3C}g_1,\\[2mm]
&d(e^{3(A+C}\sin\alpha \cos\beta)+2 m e^{2A+3C}(\cos\alpha \cos\beta V-\sin\beta W)=e^{3(A+C)}(\star_5 g_4-\cos\alpha g_1)\label{eq:nos3bps10}.
\end{align} 
\end{subequations}
These conditions give rise to two physically distinct classes of solution. To see this we first note that \eqref{eq:nos3bps1} implies we cannot set $\cos\beta=0$, however there is no barrier to fixing $\sin\beta =0$ or $\cos\alpha=0$. Taking $d$\eqref{eq:nos3bps4} and substituting \eqref{eq:nos3bps3}-\eqref{eq:nos3bps4}, and taking \eqref{eq:nos3bps2}$\wedge (\cos\alpha \cos\beta V-\sin\beta W)$ then substituting \eqref{eq:nos3bps1} we find
\beq
d(\left(\frac{\cos\alpha}{e^{3A}\sin^2\alpha}\right)\wedge W\wedge V=d(e^{2A}\sin^2\beta\sin\alpha\cos\beta \sin\beta)\wedge (\cos\alpha \cos\beta V-\sin\beta W) =0.
\eeq
These conditions, together with \eqref{eq:nos3bps1}, \eqref{eq:nos3bps6} imply that either $\sin\beta=0$ or  $(e^A,e^C,\alpha,\beta)$ are independent of the coordinates that span $\mu^i$. This is a branching of possible classes of solutions, we study these two cases in sections \ref{eq: R4class} and \ref{eq: T3class} respectively. Fixing $\cos\alpha=0$, while possible, does not lead to additional classes because it is possible to solve the above conditions without assuming $\cos\alpha\neq 0$ in both cases.\\
~\\
In the next section  we present the necessary conditions for the case with a necessary $\tilde{\text{S}}^3$ in M$_5$.

\subsubsection{Case with necessary extra 3-sphere}
The cases for which M$_5$ contains a 3-sphere necessarily require that
\beq
\sin\beta\neq0,~~~~d(\rho_1,\rho_i) \neq 0,~~~~ e^C m-e^A \cos\beta \sin\alpha \neq 0\label{eq:notzero}.
\eeq
We then have that $\text{M}_5$ decomposes as
\beq
ds^2(\text{M}_5)= \frac{e^{2(A+C)}\sin^2\alpha\sin^2\beta}{(e^C m-e^A \cos\beta \sin\alpha)^2}ds^2(\tilde{\text{S}}^3)+ U^2+V^2.
\eeq
As  everything in \eqref{eq:SU3forms} is turned on the initial $d=5$ conditions one extracts form  \eqref{BPS-conditionscomplex1}-\eqref{BPS-conditionscomplex7} are rather long winded, however with some work it becomes  apparent that the isometries of $\tilde{\text{S}}^3$ are respect by $(e^A,e^C,\alpha,\beta,g_1,g_4)$ at which point simplifying the system becomes more straight forward - having so many function we know cannot vanish helps considerably in this process because we know we can divide by each of them in expressions. After a long calculation we find that for this case sufficient conditions for supersymmetry are given by
\begin{subequations}
\begin{align}
&d(e^{A-C}\sin\alpha \cos\beta)=0,\label{eq:bpsextras31}\\[2mm]
&d(e^{2A+C}\sin\alpha \sin\beta)+ e^A(m e^{C}\cos\beta-e^A\sin\alpha)W+ m e^{A+C}\sin\beta \cos\alpha V=0,\label{eq:bpsextras32}\\[2mm]
&d(e^{2A}\sin\alpha V)=0\label{eq:bpsextras33},\\[2mm]
&d\left(\frac{1}{e^C \sin\beta} W\right)=0\label{eq:bpsextras34},\\[2mm]
& d\left(\frac{e^{A+2C}}{m e^C-e^{A}\sin\alpha \cos\beta}(\cos\beta V-\cos\alpha \sin\beta W)\right)=0,\label{eq:bpsextras35}\\[2mm]
& e^{3A} G_1 + 2 e^{2A} m \sin \alpha V - d \left(e^{3A} \cos \alpha \right) = 0,\label{eq:bpsextras35b}\\[2mm]
&d(e^{2A+3C}(\cos\alpha \cos\beta  V-\sin\beta W))=e^{2A+3C}\sin\alpha g_1\wedge V,\label{eq:bpsextras36}\\[2mm]
&d(e^{2A+3C}(\cos\alpha \cos\beta  W+\sin\beta V))+ e^{A+2C}(m e^{C}\sin\alpha\cos\beta- 3 e^A)W\wedge V= e^{2A+3C}\sin\alpha g_1\wedge W,\label{eq:bpsextras37}\\[2mm]
&d(e^{3(A+C)}\sin\alpha \cos\beta)+2 m e^{2A+3C}(\cos\alpha \cos\beta W-\sin\beta V)=e^{3(A+C)}(\star_5 g_4-\cos\alpha g_1)\label{eq:bpsextras38}.
\end{align}
\end{subequations} 
These equations are consistent with a further 2 physically distinct classes of solution, the easiest way to see this is to look at the Killing vector bilinear \eqref{eq:Killingmatrices}. Clearly the above conditions are compatible with fixing either $\cos\beta=0$ or $ \cos\alpha=0$, but making the former choice makes $k^{\pm}_i$ drop out of $K^{IJ\pm}$ which suggests that the original S$^3$ is not actually an R-symmetry, but rather an outer automorphism of the algebra. One can also see this branching of possibilities from \eqref{eq:bpsextras31}-\eqref{eq:bpsextras34}: One solves \eqref{eq:bpsextras33}-\eqref{eq:bpsextras34} in general in terms of two local coordinates and uses \eqref{eq:bpsextras31} to define $\cos\beta$ in terms of a constant, \eqref{eq:bpsextras32} then gives two PDEs involving $(e^A,e^C,\alpha)$ and this constant, but solving them requires you to decide whether $\cos\beta=0$, or not, which leads to two different classes. We study these in section \ref{eq:wickrotatedclass} and \ref{largeclass} respectively, only the latter class with $\cos\beta\neq 0$ supports the large ${\cal N}=(4,4)$, algebra in line with our expectations from considering \eqref{eq:Killingmatrices}. 
\section{3 Classes of solutions with small ${\cal N}=(4,4)$}\label{eq:smallclasses}
In this section we derive and present all classes of supporting the small ${\cal N}=(4,4)$ superconformal algebra, there are 3 of them, though only 2 are independent.

\subsection{AdS$_3 \times $ S$^3 \times \mathbb{R}^4 \times \mathbb{R}^1 $ warped product }\label{eq: R4class}
In this section we derive the class of solutions that results from imposing $\sin\beta=0$ in the conditions \eqref{eq:nos3bps1}-\eqref{eq:nos3bps10}, as we shall see the result is a generalisation of the near horizon limit of a certain brane intersection considered in \cite{Dibitetto:2020bsh}.\\
~\\
We can without loss of generality set $\sin \beta = 0$ by taking $\beta=0$ upon which the system of conditions in \eqref{eq:nos3bps1}-\eqref{eq:nos3bps7} can be expressed as 
\begin{subequations}
\begin{align}
& me^C- e^A \sin\alpha=0,\label{eq:small1bps1}\\[2mm]
& d(e^{2A}\sin\alpha V)=0,\label{eq:small1bps2}\\[2mm]
& d(e^{2A}\sin\alpha W)+ e^{A}\cos\alpha m V\wedge W=d(e^{2A}\sin\alpha \mu^i)+ e^{A}\cos\alpha m V\wedge \mu^i=0,\label{eq:small1bps3}\\[2mm]
&\left(d\left(\frac{\cos\alpha}{e^{3A}\sin^2\alpha}\right)- 2m \frac{\cos^2\alpha}{e^{4A}\sin^3\alpha}V\right)\wedge W=0\label{eq:small1bps4},
\end{align}
\end{subequations}
while \eqref{eq:nos3bps8}-\eqref{eq:nos3bps10} just serve as definitions of the flux. We take \eqref{eq:small1bps1} to define $e^C$ and solve \eqref{eq:small1bps2} by introducing a local coordinate $\rho$ such that 
\beq
e^{2A}\sin\alpha V= -\frac{1}{m}d\rho,
\eeq
we then find from $d$\eqref{eq:small1bps3} that 
\beq
d\left(\frac{\cos\alpha}{e^{3A}\sin^2\alpha}\right)\wedge d\rho\wedge \mu^i=0.
\eeq
Since $\mu^i$ define independent vielbein direction orthogonal to $d\rho$ this implies that the function under $d$ has support on $\rho$ alone, so we can solve this condition in terms of a local function $u=u(\rho)$ as
\beq
\frac{\cos\alpha}{e^{3A}\sin^2\alpha}=- \frac{1}{2}\partial_{\rho}\log u.
\eeq
This allows us to express \eqref{eq:small1bps3} as
\beq
d(e^{2A} u^{-\frac{1}{2}}\sin\alpha \mu^i)=d(e^{2A} u^{-\frac{1}{2}}\sin\alpha W)=0.
\eeq
We solve these conditions in terms of 4 local coordinates $x_1,...x_4$ as
\beq
e^{2A} u^{-\frac{1}{2}}\sin\alpha \mu^i= \frac{1}{m}dx_{i},~~~~e^{2A} u^{-\frac{1}{2}}\sin\alpha W= \frac{1}{m} dx_{4},
\eeq
such that M$_4$ is conformally $\mathbb{R}^4$. The final purely geometric condition \eqref{eq:small1bps4} then reduces to simply
\beq
\partial_{\rho}^2u=0,
\eeq
imposing that $u$ is globally a linear function. If we then take \eqref{eq:nos3bps8}-\eqref{eq:nos3bps10} to define $(G_1,g_1,g_4)$ the conditions for ${\cal N}=(4,4)$ supersymmetry are solved, however we find it convenient to introduce a final function $h=h(\rho,x_1,...,x_4)$  in terms of which
\beq
e^{6A}\sin^2\alpha= \frac{u^2}{h},
\eeq
before presenting the class we have derived.\\
~\\
In summary our first class of solutions, in $d=11$ notation for the flux, can be expressed as
\begin{align}
ds^2 &=  \Delta_1^{\frac{1}{3}} \bigg[\frac{u^{\frac{2}{3}}}{h^{\frac{1}{3}}} \left( ds^2(\text{AdS}_3) + \frac{1}{m^2 \Delta_1} ds^2(\text{S}^3) \right)+ \frac{1}{m^2}\frac{h^{\frac{2}{3}}}{u^{\frac{4}{3}}} \left(d\rho^2+ u ds^2(\mathbb{R}^4) \right)\bigg],\nn \\[2mm]
G &=  d \left( \frac{u u'}{2  h}+2 \rho  \right)\wedge \text{vol}(\text{AdS}_3)+ 
\frac{1}{m^3} d\left( \frac{u u'}{2  h \Delta_1}-2\rho  \right)\wedge \text{vol}(\text{S}^3) \label{eq:smallclassI} \\[2mm]
& + \frac{1}{m^3}\left(\frac{1}{u} \star_4 d_4 h \wedge d\rho+ \partial_{\rho}h \text{vol}(\mathbb{R}^4)\right), \qquad \Delta_1 =1+\frac{u'^2}{4 h} . \nn
\end{align}
Where $u$ is a linear function of $\rho$, while $h$ has support on $(\rho,\mathbb{R}^4)$ and $\star_4$ is the hodge dual on unwarped $\mathbb{R}^4$. \footnote{In terms of the coordinates $x_{a}$ for $a=1,...,4$ we have
\beq
ds^2(\mathbb{R}^4)=(dx_a)^2,~~~~~\star_4 d_4h= \frac{1}{3!}\epsilon_{abcd}\partial_{x_a} h \, dx_b\wedge dx_c\wedge dx_d.\nn
\eeq}
The first two terms in $G$ are manifestly closed, but not the part defined  strictly within $(\rho,\mathbb{R}^4)$. Away from the loci of sources this should also be closed which leads to an additional PDE 
\begin{align}
\frac{1}{u}\nabla^2_{\mathbb{R}^4} h +  \partial_{\rho}^2 h =0,\label{eq:smallclass1pde}
\end{align}
where the right hand of this expression should be modified at the loci of sources - this equation implies \eqref{magnetic} and so all the equations of motion follow.  Solutions in this class are then in one to one correspondence with solutions to $\partial_{\rho}^2u=0$ and \eqref{eq:smallclass1pde}, which is a deformation of the PDE one expects an M5 brane in flat space to obey, reducing to it when $u=1$.\\
~~\\
Given that this class generically only has an SO$(2,2)\times$SO$(4)$ it should be obvious that, lacking the required R-symmetry, it does not experience an enhancement to large ${\cal N}=(4,4)$. Indeed computing the 1-forms dual to the canonical Killing vectors the spinors defines for this case we find
\beq
{\cal K}^{IJ\pm}= \frac{c m}{2} e^{2C} k^{\pm}_i
\eeq
which contain only the SU(2)$_{\pm}$ charged forms on S$^3$, consistent with our expectations.

\subsubsection{Near horizon limit}
\label{Nearhorizon}
We note that if we decompose $\mathbb{R}^4$ in spherical polar coordinates and demand that $h$ is a singlet with respect to the isometries of resulting $\tilde{\text{S}}^3$ it is possible to map our solution to the class of small ${\cal N}=(4,4)$ solutions found as a near horizon limit of a stack of M$5$ branes, a stack of M$2$ branes and additional dyonic M$5$ branes found in \cite{Dibitetto:2020bsh}. Indeed the entire class we derive here can be realised as a near horizon limit with a slight modification of their ansatz. Instead of assuming there are two S$^3$, we will work with a general $\mathbb{R}^4$. The brane picture becomes
\begin{table}[h!]
\renewcommand{\arraystretch}{1}
\begin{center}
\scalebox{1}[1]{
\begin{tabular}{c||c c|c c c c|c c c c |c}
branes & $t$ & $x$  & $r$ & $\theta^{1}$ & $\theta^{2}$ & $\theta^{3}$ & $x_1$ & $x_2$ & $x_3$ & $x_4$ & $z$ \\
\hline \hline
$\mathrm{M5}_{\xi}$ & $\times$ & $\times$  & $\times$ & $\times$ & $\times$ & $\times$ & $-$ & $-$ & $-$ & $-$ & $-$  \\
\hline
$\mathrm{M2}$ & $\times$ & $\times$ & $-$ & $-$ & $-$ & $-$ & $\sim$ & $\sim$ & $\sim$ & $\sim$ & $\times$ \\
$\mathrm{M5} $ & $\times$ & $\times$ & $-$ & $-$ & $-$ & $-$ & $\times$ & $\times$ & $\times$ & $\times$ & $\sim$ \\
\end{tabular}
}
\end{center}
\caption{{Brane picture describing the intersection of $\mathrm{M5_{\xi}}$-$\mathrm{M5}$-$\mathrm{M2}$ branes}} \label{Table:Mbranes}
\end{table}
while the metric and the 4-form are generalised to 
\begin{align}
\label{lambertdefect}
ds^{2} & = \frac{f^{2/3} \left(s^2+c^2H\right)^{1/3}}{H^{2/3}}\left( \frac{1}{H_{\mathrm{M}_2}^{2/3}H_{\mathrm{M}_5}^{1/3}} ds^2(\text{Mink}_2) + H_{\mathrm{M}_2}^{1/3}H_{\mathrm{M}_5}^{2/3} dr^2\right)\nn \\[1.5mm]
& + \frac{f^{2/3} \, H^{1/3}}{\left(s^2+c^2H\right)^{2/3}} H_{\mathrm{M}_2}^{1/3}H_{\mathrm{M}_5}^{2/3} \delta^2 r^2 ds^2(\text{S}^3) + \frac{H^{1/3}\,\left(s^2+c^2H\right)^{1/3}}{f^{4/3}}\,\left(  \frac{H_{\mathrm{M}_5}^{2/3}}{H_{\mathrm{M}_2}^{2/3}} dz^2 + f \frac{H_{\mathrm{M}_2}^{1/3}}{H_{\mathrm{M}_5}^{1/3}}  ds^2(\mathbb{R}^4) \right), \nn \\[1.5mm]
G & = - c \, d\left(\frac{1}{H_{\mathrm{M}_2}}\right)\wedge \text{vol}(\text{Mink}_2) \wedge dz + \frac{1}{c^2} \star_{4_{\delta}} d H_{\mathrm{M}_5} \wedge dz -\frac{s}{\sqrt{H_{\mathrm{M}_2}}} \text{vol}(\text{Mink}_2)\wedge dr \wedge d\left(\frac{f}{H}\right) \nn \\[1.5mm]
& - r^3  s \, c^2 \, H_{\mathrm{M}_5}^{3/2} \text{vol}(\text{S}^3) \wedge d\left(\frac{f}{c^2 H + s^2}\right)+ c \left(\frac{1 }{f}  \star_4 dH \wedge dz - \frac{ H_{\mathrm{M}_2}}{H_{\mathrm{M}_5}}  \partial_z H \text{vol}(\mathbb{R}_4) \right),
\end{align}
where $c = \cos \xi$ and $s = \sin \xi$, $\delta$ is the deficit angle of $\mathbb{R}^4_{\delta}$ and $\xi$ is the deficit angle of M$5_{\xi}$. $ \star_{4_{\delta}}$ is operated along $\mathbb{R}^4_{\delta}$, $\star_4$ along $\mathbb{R}^4$, $f$ is a linear function of $z$, $f'' = 0$ and $H$ satisfies the modified Laplace equation
\begin{align}
\frac{1}{f} \Delta_{\mathbb{R}^4} H + \partial_z^2 H = 0.
\end{align}
Solving the EOM imposes
\begin{align}
\delta = \cos \xi, \qquad H_{\mathrm{M}_2} = H_{\mathrm{M}_5} = \frac{Q}{r^2}, \qquad f'= \frac{2}{\sqrt{Q}} \tan \xi.
\end{align}
If we plug these definitions in \eqref{lambertdefect} we get \eqref{eq:smallclassI} upon identyfing
\begin{align}
\label{eq:coordtrans}
f = \frac{2 \,  u \tan \xi \sec \xi}{m^3 \, Q^{3/2} \, u' } , \qquad H = \frac{4 \, h \tan ^2 \xi }{u'^2}, \qquad z \to \frac{\sec \xi }{m^3 \, Q^{3/2}} z, \qquad x_i \to \frac{u' \csc^{1/2} \xi }{\sqrt{2} \, m^{3/2} \, Q^{1/4}}.
\end{align}
This is well defined in the case $u' \neq 0$. When $u$ is constant we define $ u' = \sin \xi$ then we send $ \xi \to 0 $, in this way \eqref{eq:coordtrans} still holds.

\subsection{Foliation of AdS$_3 \times $ S$^3 \times \mathbb{T}^3$ over $ \Sigma_2$}\label{eq: T3class}
Our next class of solutions follows from assuming $\sin\beta\neq 0$ in the conditions \eqref{eq:nos3bps1}-\eqref{eq:nos3bps10}, this leads to a class with a $\mathbb{T}^3$ in the internal space whose isometries are respected.\\
~\\
We have already shown that imposing $\sin\beta \neq 0$ means that $(e^{A},e^C,\alpha,\beta)$ are independent of the coordinates spanning $\mu^i$ and that $(\cos\beta\neq0,\sin\beta\neq 0,\sin\alpha\neq0)$. Given this we can further simplify the conditions \eqref{eq:nos3bps1}-\eqref{eq:nos3bps7}, we find they reduce to
\begin{subequations}
\begin{align}
&me^{C}-e^{A}\sin\alpha\cos\beta=0,\label{eq:BPSclassIIsmall1}\\[2mm]
&d(e^{2A+C}\sin\alpha \sin\beta)+e^{2A}\sin\alpha \sin\beta(\cos\alpha\cos\beta V-\sin\beta W )=0,\label{eq:BPSclassIIsmall2}\\[2mm]
&d(e^{2A}\sin\alpha \sin\beta(\cos\alpha \cos\beta W+\sin\beta V)=0,\label{eq:BPSclassIIsmall3}\\[2mm]
&d\left(\frac{1}{e^A \sin\alpha \sin\beta} \mu^i\right)=0,\label{eq:BPSclassIIsmall4}\\[2mm]
&d(e^{2A}\sin\alpha V)=0,\label{eq:BPSclassIIsmall5}\\[2mm]
&d\left(\frac{1}{e^{A}\sin\alpha \sin\beta}(\cos\beta W+\cos\alpha\sin\beta V)\right)=0,\label{eq:BPSclassIIsmall6}
\end{align}
\end{subequations}
where \eqref{eq:BPSclassIIsmall1} merely defines $e^C$ and the fluxes are still defined by \eqref{eq:nos3bps8}-\eqref{eq:nos3bps10}. The first thing we need to do is choose which of the 1- and 2-form constraints we use to define a vielbein, there are several options but we note that using \eqref{eq:BPSclassIIsmall2}, \eqref{eq:BPSclassIIsmall3} leads to a Riemann surface with common warping - we thus solve \eqref{eq:BPSclassIIsmall2}-\eqref{eq:BPSclassIIsmall3} in terms of the local coordinates $(\rho,x,z_i)$ as
\begin{align}
&e^{2A}\sin\alpha \sin\beta(\cos\alpha\cos\beta V-\sin\beta W )=-\frac{L^3}{m}d\rho,\nn\\[2mm]
&e^{2A}\sin\alpha \sin\beta(\cos\alpha \cos\beta W+\sin\beta V)=\frac{L^3}{m}dx,\label{eq:localcoordsclass2}\\[2mm]
&\frac{1}{e^A \sin\alpha \sin\beta} \mu^i= \frac{1}{m}dz_i\nn,
\end{align}
where $\partial_{z_i}(e^{A},e^C,\alpha,\beta)=0$ and we see that $\mu^i$ span either $\mathbb{R}^3$ or $\mathbb{T}^3$, we will assume the latter.   Consistency with \eqref{eq:BPSclassIIsmall2} then demands that we impose
\beq
e^{3A}\sin^2\alpha \sin\beta \cos\beta=L^3\rho.
\eeq
We are left with \eqref{eq:BPSclassIIsmall5} and \eqref{eq:BPSclassIIsmall6} to deal with which can now be expressed as
\begin{subequations}
\begin{align}
d\left(\frac{\cos\alpha\cot\beta dx+\sin^2\alpha\cos^2\beta d\rho}{\rho(\cos^2\alpha\cos^2\beta+\sin^2\beta)}\right)=0,\label{eq:locallyclosed1}\\[2mm]
d\left(\frac{dx-\cos\alpha \cot\beta d\rho}{\cos^2\alpha\cos^2\beta+\sin^2\beta}\right)=0,\label{eq:locallyclosed2}
\end{align}
\end{subequations}
meaning the quantities under the exterior derivative are locally closed. 
We solve \eqref{eq:locallyclosed1} by redefining $(\alpha,\beta)$ in terms of a local function $h=h(\rho,x)$ such that\footnote{There are some arbitrary signs that enter when solving the following condition, we fix them such that $(\sin\alpha,\cos\alpha,\sin\beta,\cos\beta)$ all come a $+$ sign without loss of generality.}
\beq
\frac{\cos\alpha\cot\beta dx+\sin^2\alpha\cos^2\beta d\rho}{\rho(\cos^2\alpha\cos^2\beta+\sin^2\beta)}= d (h-\log\rho),
\eeq
upon which \eqref{eq:locallyclosed2} reduces to 
\beq
\left(\frac{1}{\rho}\partial_{\rho}(\rho \partial_{\rho}h)+\partial_x^2h\right)\rho d\rho\wedge dx=0,\label{eq:class2bpspde}
\eeq
which means that $h$ must satisfy a  Laplace equation in 3 dimensions in cylindrical polar coordinate with axial symmetry. The fluxes $(G_1,g_1,g_4)$ can now be straight forwardly extracted from \eqref{eq:nos3bps8}-\eqref{eq:nos3bps10} upon which the conditions for ${\cal N}=(4,4)$ have reduced to a single condition \eqref{eq:class2bpspde}.\\
~~\\
In summary this class of solutions can be written as\footnote{In terms of the local coordinates introduced in \eqref{eq:localcoordsclass2} we have
\beq
ds^2(\mathbb{T}^3)=\sum_i dz_i,~~~~\text{vol}(\mathbb{T}^3)=dz_1\wedge dz_2\wedge dz_3\nn
\eeq}
\begin{align}\label{eq:smallclassII}
\frac{ds^2}{L^2}&=\rho^{\frac{2}{3}}\left(\frac{\Lambda_1\Lambda_2}{\Lambda^2_3}\right)^{\frac{1}{3}}ds^2(\text{AdS}_3)+\frac{\rho^{\frac{2}{3}}}{m^2} \bigg[\left(\frac{\Lambda_1\Lambda_3}{\Lambda^2_2}\right)^{\frac{1}{3}}ds^2(\text{S}^3)+\left(\frac{\Lambda_2\Lambda_3}{\Lambda^2_1}\right)^{\frac{1}{3}}ds^2(\mathbb{T}^3) \nn \\[2mm]
& + \Lambda_0 (\Lambda_1\Lambda_2\Lambda_3)^{\frac{1}{3}}(d\rho^2+dx^2)\bigg], \\[2mm]
\frac{G}{L^3}&=\bigg(d\left(\frac{\partial_x h}{\Lambda_0 } \frac{1}{\Lambda_3} \right)+2 \rho(\partial_x hd\rho- \partial_{\rho}h dx)\bigg)\wedge \text{vol}(\text{AdS}_3) \nn \\[2mm]
&+\frac{1}{m^3}\bigg(d\left(\frac{\partial_x h}{ \Lambda_0 } \frac{1}{\Lambda_2} -2x\right)-2 \rho(\partial_x hd\rho- \partial_{\rho}h dx)\bigg)\wedge \text{vol}(\text{S}^3) + \frac{1}{m^3}d\left(\frac{\partial_x h}{ \Lambda_0 } \frac{1}{\Lambda_1} \right)\wedge\text{vol}(\mathbb{T}^3), \nn
\end{align}
where we introduced $\Lambda_0$ to make the notation more compact
\begin{align}
\Lambda_0 = (\partial_{x}h)^2 + (\partial_{\rho}h)^2,
\end{align}
and $\Lambda_{1,2,3}$ are functions defined as 
\begin{equation}
\begin{split}
\Lambda_1  = 1 - \frac{\partial_{\rho} h }{\rho \Lambda_0 }, ~~~ \Lambda_2 = \frac{\partial_{\rho} h }{\rho \Lambda_0 }, ~~~ \Lambda_3  = -\frac{ 1 - \rho \partial_{\rho} h }{\rho^2 \Lambda_0 } . \label{eq:Lambdassmallclass}
\end{split}
\end{equation}
Solutions in this class are governed by a cylindrical Laplace equation
\beq
\frac{1}{\rho}\partial_{\rho}(\rho \partial_{\rho}h)+ \partial_{x}^2h =0.
\eeq
We find that the canonical matrix 1-forms of \eqref{eq:Killingmatrices} become
\begin{equation}
\begin{split}
{\cal K}^{+}&= c m\left(\frac{e^{2C}}{2}\cos\beta k^{+}_i\Sigma^i+e^{2D}dz_i\tilde{\Sigma}^i\right),~~~~{\cal K}^{-}=  c m\left(\frac{e^{2C}}{2}\cos\beta k^{-}_i\Sigma^i+e^{2D}dz_i\tilde{\Sigma}^i\right), \\[2mm]
e^{2C}&=\frac{\rho^{\frac{2}{3}}}{m^2}\left(\frac{\Lambda_1\Lambda_3}{\Lambda^2_2}\right)^{\frac{1}{3}},~~~~e^{2D}=\frac{\rho^{\frac{2}{3}}}{m^2}\left(\frac{\Lambda_2\Lambda_3}{\Lambda^2_1}\right)^{\frac{1}{3}},
\end{split}
\end{equation}
where $(e^{2C},e^{2D})$ are the warp factors of S$^3$ and $\mathbb{T}^3$ respectively, making it clear the vectors dual to this are indeed Killing. It is a simple matter to show that the internal $d=8$ spinors are singlets with respect to $\partial_{z_i}$, as such there is no enhancement of the R-symmetry which remains  SU(2)$_{+}\times $ SU(2)$_{-}$.
\subsection{Foliation of AdS$_3 \times $ S$^3 \times \tilde{\text{S}}^3$ over  $\Sigma_2$ }\label{eq:wickrotatedclass}
Looking at \eqref{eq:bpsextras31}-\eqref{eq:bpsextras38} we notice that we need to discern two different cases according to how we solve the differential constraints: $\cos \beta = 0$ or $\cos \beta \neq 0$. In this section we focus on the first case. This gives rise to a class that preserves small ${\cal{N}}=(4,4)$ supersymmetry. \eqref{eq:bpsextras31}- \eqref{eq:bpsextras35} can be arranged to be
\begin{subequations}
\begin{align}
& d(e^{2A+C}\sin\alpha) - e^{2A}\sin\alpha W+ m e^{A+C}  \cos\alpha V=0,\label{eq:BPSclassIIIsmall1}\\[2mm]
& d(e^{2A}\sin\alpha V)=0,\label{eq:BPSclassIIIsmall2} \\[2mm]
& d\left(e^{-C} W\right)=0, \label{eq:BPSclassIIIsmall3} \\[2mm]
& d\left(e^{A+C} \cos\alpha \right) \wedge W =0. \label{eq:BPSclassIIIsmall}
\end{align}
\end{subequations}
We use \eqref{eq:BPSclassIIIsmall2} and \eqref{eq:BPSclassIIIsmall3} to define the Riemann surface
\begin{align}
e^{2 A} \sin \alpha V = - \frac{1}{m} d\rho , \qquad e^{-C}W=\frac{dr}{r}.
\end{align}
We  now find it convenient to introduce two new functions $u$ and $g$ that are independent and must respect the S$^3$ isometries, so depend only on $r$ and $\rho$. These functions are defined as
\begin{align}
& e^{2 A + C} \sin \alpha = \frac{r}{m} \sqrt{u}, \qquad e^{A + 2 C} \cos \alpha = r g.
\end{align}
where $u = u(r,\rho)$ and $g = g(r,\rho)$. The conditions \eqref{eq:BPSclassIIIsmall1} and \eqref{eq:BPSclassIIIsmall} become
\begin{align}
- 2 m g + r \partial_{\rho} u = 0, \qquad \partial_r u =0, \qquad d g \wedge dr =0.
\end{align}
These imply
\begin{align}
u = u(\rho), \qquad u'' =0,
\end{align}
and
\begin{align}
g = \frac{r u' }{2 m}.
\end{align}
The flux equations \eqref{eq:bpsextras35b}-\eqref{eq:bpsextras38} define $(g_1,g_4)$ in a self consistent way, but we find it useful to introduce a new function arbitrary $h=h(r,\rho)$ such that
\begin{align}
e^{6A} \sin^2 \alpha = \frac{u^2}{h},
\end{align}
before presenting them along with a summary of the class.
~\\
In summary this class takes the form
\begin{equation}
\begin{split}\label{eq:class3small}
ds^2&=\Delta_1^{\frac{1}{3}}\bigg[\frac{u^{\frac{2}{3}}}{h^{\frac{1}{3}}}\left(ds^2(\text{AdS}_3)+\frac{1}{m^2\Delta_1}ds^2(\tilde{\text{S}}^3)\right)+ \frac{1}{m^2}\frac{h^{\frac{2}{3}}}{u^{\frac{4}{3}}}\bigg(d\rho^2+ u\left(dr^2+r^2 ds^2(\text{S}^3))\right)\bigg)\bigg], \\[2mm]
G&= d\left(\frac{u u'}{2 h} + 2\rho \right)\wedge \text{vol}(\text{AdS}_3)-\frac{r^3}{m^3}\left(\frac{1}{u}\partial_{r}h d\rho-\partial_{\rho}hdr\right)\wedge \text{vol}(\text{S}^3) \\[2mm]
& + \frac{1}{m^3}d\left(\frac{u u'}{2 h \Delta_1} - 2\rho \right)\wedge \text{vol}(\tilde{\text{S}}^3), \\[2mm]
\end{split}
\end{equation}
where $u$ is a linear function of $\rho$, $h$ has support on $(\rho,r)$ and $\Delta_1$ is defined as in \eqref{eq:smallclassI}. Solutions in this class are governed by $\partial_{\rho}^2u=0$ and the PDE following from the Bianchi identity of $G$:
\beq
\frac{1}{u}\frac{1}{r^3}\partial_{r}(r^3 \partial_r h)+\partial_{\rho}^2h=0, \\[2mm]
\eeq
when these are solved on has a solution. The above PDE is simply that of  \eqref{eq:smallclass1pde} with SO(4) isometry imposed on the $\mathbb{R}^4$ Laplacian there. Indeed it should not be hard to see that the entire class of this section is a subclass that of section \ref{eq: R4class}.

Although this class is merely a subclass of section \ref{eq: R4class} it does illiterate one interesting point, namely that assuming a 3-sphere does not necessarily imply that it is realising an R-symmetry. Clearly S$^3$, which we assumed the presence of from the start, is actually appearing as part of $\mathbb{R}^4$ expressed in polar coordinates in \eqref{eq:class3small}. This is indicative of the SO(4) of S$^3$ not realising an R-symmetry, despite the spinors being charged  under this group by construction. A similar scenario also exists in D1-D5 near horizon geometry AdS$_3\times$S$^3\times \mathbb{T}^4$. Here the $\mathbb{T}^4$ is locally $\mathbb{R}^4$ and while the spinors of this solution are not charged under the U(1)$^4$ of $\mathbb{T}^4$ they are charged under the SO(4) of this $\mathbb{R}^4$. In this case this SO(4) actually realises an outer automorphism of the small superconformal algebra, and the same thing is happening in our class here. We also see this if we compute the 1-forms dual to the canonical Killing vectors, finding in this case
\begin{align}
{\cal{K}}^{\pm} = - \frac{c m }{2} e^{2 \tilde{C}} \tilde{k}^{\pm}_i \tilde{\Sigma}^i,
\end{align}
where $e^{\tilde{C}}$ is the warp factor of $ds^2(\tilde{\text{S}}^3)$
\begin{align}
e^{\tilde{C}} = \frac{u^{\frac{1}{3}}}{m \, h^{1/6} \Delta_1^{1/3} },
\end{align}
which indicates that it is $\tilde{k}^{\pm}_i$ and not $k^{\pm}_i$ that realises the SU(2)$\times$SU(2)  R-symmetry.

\section{Class of solutions with large ${\cal{N}}=(4,4)$ }\label{largeclass}
In this section we present the only class of solutions that preserve large ${\cal{N}}=(4,4)$. \\
~\\
In this class we assumed that $\cos \beta \neq 0$, this does not simplify anymore the conditions \eqref{eq:bpsextras31}-\eqref{eq:bpsextras38}, where we need \eqref{eq:bpsextras35b}-\eqref{eq:bpsextras38} only to define the fluxes. We start by solving \eqref{eq:bpsextras31}-\eqref{eq:bpsextras35}. We can use the two relations that define the vielbein on the Riemann surface \eqref{eq:bpsextras33} and \eqref{eq:bpsextras34} and write them as
\begin{equation}
\begin{split}
 e^{2 A} \sin \alpha V = - \frac{L^3}{2 m} dy ,~~~ \frac{W}{e^C \sin \beta }= -\frac{\gamma}{2} dr.
\end{split}
\end{equation}
At this point we turn our attention to \eqref{eq:bpsextras31}, which we can solve in terms of a constant $\gamma$ as
\begin{align}
 e^{A-C}  \sin \alpha \cos \beta  = \gamma m.
\end{align}
This gives us a definition for $e^C$, but note that one cannot fix $\gamma=0,1$ without setting the warp factors of S$^3$ or $\tilde{\text{S}}^3$ to either $0$ or $\infty$. We can plug this definition along with the expressions for $V$ and $W$ into \eqref{eq:bpsextras32}. Locally, we can write the first term as a function that depends on the two coordinates on the Riemann surface $y, r$ times an overall constant
\begin{align}
e^{2 A + C } \sin \alpha \sin \beta = \frac{L^3}{\gamma m} e^{\frac{1}{2} D} .
\end{align}
yeilding a definition for $e^A$. Thanks to the function $D$ we have just introduced and using the local coordinates \eqref{eq:bpsextras32} becomes
\begin{align}
& 2 \, d\left(e^\frac{D}{2}\right)- \cos \alpha \sin \beta \cos \beta dy =  e^{\frac{D}{2}} \left( \cos ^2 \beta -\gamma \right) dr.
\end{align}
It provides two different equations along $dy$ and $dr$ that allow us to define $\cos \alpha$ and $\cos \beta$ in terms of $D$ and its derivatives. We rewrite \eqref{eq:bpsextras35} using all the ingredients we have as
\begin{align}
d \left(  \left(\partial_r D + \gamma \right) dy - e^{D} \partial_{y} D \, dr \right)=0.
\end{align}
It is solved if the function $D$ satisfy the following toda equation
\begin{align}
\partial_r^2 D + \partial_y^2 e^D = 0.
\end{align}
All the purely geometric conditions for supersymmetry are now solved and we can extract the fluxes from \eqref{eq:bpsextras35b}- \eqref{eq:bpsextras38}. \\
~~\\
In summary the metric and the flux for this class can be expressed as
\begin{align}
\label{todalarge}
\frac{ds^2}{L^2 } & = e^{\frac{D}{3}}\biggl( \left( \frac{\Lambda_1 \Lambda_2}{\Lambda_3^2} \right)^{1/3}ds^2(\text{AdS}_3) + \frac{1}{m^2} \biggl[ \frac{1}{\gamma^2} \left( \frac{\Lambda_1 \Lambda_3}{\Lambda_2^2} \right)^{1/3} ds^2(\text{S}^3) + \frac{1}{(1-\gamma)^2} \left( \frac{\Lambda_2 \Lambda_3}{\Lambda_1^2} \right)^{1/3} ds^2(\tilde{\text{S}}^3) \nn \\[2mm]
& + \frac{(\Lambda_1 \Lambda_2 \Lambda_3)^{1/3}}{4} \left( dr^2 + e^{- D} dy^2 \right)\biggr] \biggr), \nn \\[2mm]
\frac{G}{L^3} & =  d\left(  \frac{e^D \partial_y D}{ \Lambda_3 } + y \right)\wedge \text{vol}(\text{AdS}_3) \\[2mm]
& + \frac{1}{m^3} \left( d\left( \frac{e^D \partial_y D}{\gamma^3 \Lambda_2 } - \frac{y}{\gamma} \right)  - \frac{1}{\gamma^2} \left( \partial_r D ~ dy - e^D  \partial_y D ~ dr  \right) \right) \wedge \text{vol}(\text{S}^3) \nn \\[2mm]
& + \frac{1}{m^3} \left( d \left( \frac{e^D \partial_y D}{(1-\gamma)^3 \Lambda_1 } - \frac{y}{1-\gamma} \right) + \frac{1}{(1-\gamma)^2} \left( \partial_r D ~ dy - e^D  \partial_y D ~ dr  \right) \right) \wedge \text{vol}(\tilde{\text{S}}^3).\nn
\end{align}
The three functions $\Lambda_i$ are
\begin{align}
\Lambda_1 = \gamma + \partial_r D, ~~~ \Lambda_2 = 1 - \gamma - \partial_r D, ~~~ \Lambda_3 = (1 - \gamma - \partial_r D)(\gamma + \partial_r D) - e^D (\partial_y D)^2. \label{Lambdatoda} 
\end{align}
$D$ is a function that depends only on the coordinate on the Riemann surface $y$ and $r$ and it satisfies the toda equation
\begin{align}
\partial_r^2 D + \partial_y^2 e^D = 0.\label{eq:todaeq}
\end{align}
Parametrizing the solution in this way makes manifest the importance the parameter $\gamma$ plays.
We note that
\beq
\gamma\to 1-\gamma,~~~~r\to-r~~~~\Rightarrow ~~~~ (\Lambda_1,\Lambda_2,\Lambda_3)\to (\Lambda_2,\Lambda_1,\Lambda_3)
\eeq
leaves the Laplace equation invariant and has the effect of exchanging the two 3-spheres and sending $G\to-G$, a sign change we can evade with $L\to -L$.
This is consistent with the isomorphism of the algebra $\mathfrak{d}(2,1;\alpha)$   under $\alpha\to\frac{1}{\alpha}$ if we identify
\beq
\alpha =\frac{\gamma}{1-\gamma}.
\eeq
As described above, we can compute the one-form bilinears on the 11-dimensional background dual to the Killing vectors. In this case M$_8 = \text{S}^3 \times \tilde{\text{S}}^3 \times \Sigma_2$ and the general relation is \eqref{eq:Killingmatrices}, they are
\begin{equation}
{\cal{K}}^+ =  \frac{c m}{2} \left( \gamma e^{2 C} k^+_i \Sigma^i -  e^{2 \tilde{C}}(1 - \gamma)\tilde{k}^+_i \tilde{\Sigma}^i \right), ~~ {\cal{K}}^- = \frac{c m}{2} \left( \gamma e^{2 C} k^-_i \Sigma^i -  e^{2 \tilde{C}}(1 - \gamma) \tilde{k}^-_i \tilde{\Sigma}^i \right),
\end{equation}
where $e^{2 C}$ and $e^{2 \tilde{C}}$ are the two warp factors of the two 3-sphere, respectively of S$^3$ and $\tilde{\text{S}}^3$
\begin{align}
e^{2C} = \frac{e^{D/3}L^2}{m^2 \gamma^2} \left(\frac{\Lambda_1 \Lambda_3}{\Lambda_2^2}\right)^{1/3}, \qquad e^{2\tilde{C}} = \frac{e^{D/3}L^2}{m^2 (1-\gamma)^2} \left(\frac{\Lambda_2 \Lambda_3}{\Lambda_1^2}\right)^{1/3}. \nn
\end{align}
As we now have 4 independent sets of SU(2) invariant forms appearing in ${\cal{K}}^{\pm}$ we see that these  are consistent with large ${\cal N}=(4,4)$ - we also note, with respect to the classes preserving the small algebra, the additional dependence on the parameter $\gamma$ in this case. \\
~\\
The simplest solution in this class comes from solving \eqref{eq:todaeq} as simply $e^D=$ constant upon which the warp factors of AdS$_3$, S$^3$ and $\tilde{\text{S}}^3$ all become constants, this is of course locally the AdS$_3\times\text{S}^3\times \tilde{\text{S}}^3\times \mathbb{T}^2$ solution of \cite{Boonstra:1998yu}. If however $e^D\neq $ constant it is possible to map \eqref{eq:todaeq} to a more familiar axially symmetric Laplace equation, as we show in the next section.

\subsection{Mapping the Toda to the Laplacian}\label{eq:laplaceversion}
The class of solutions presented above preserves large ${\cal{N}}=(4,4)$, it has been written such that the PDE the function $D$ must satisfies is a Toda equation. Sadly Toda equations are famously difficult to solve, so in this section we shall map it to a more amiable cylindrical Laplace equation, at the cost of the resulting class no longer containing AdS$_3\times\text{S}^3\times \tilde{\text{S}}^3\times \mathbb{T}^2$.\\
~\\
A similar procedure has been done in \cite{Lin:2005nh}. Indeed using the same coordinate transformation
\begin{align}
& y = \rho \partial_{\rho} V, \qquad r = \partial_x V, \qquad e^{D} = \rho^2,
\end{align}
one finds that the Toda is mapped to
\beq
 \frac{1}{\rho} \partial_{\rho} (\rho \partial_{\rho} V) + \partial_x^2 V = 0,\label{eq:lap}
\eeq
an axially symmetric $d=3$ cylindrical Laplace equation. However, we observe that in the case at hand all the bosonic fields depend on $\partial_x V$ and its derivatives, and if $V$ satisfies \eqref{eq:lap} so does $\partial_x V$. We thus find it convenient to introduce a new function $h$ as
\begin{align}
h = - \frac{1}{2} \partial_x V.
\end{align}
In these new coordinates and in terms of $h$ the class of \eqref{todalarge} then becomes
\begin{align}
\frac{ds^2}{L^2}&=\rho^{\frac{2}{3}}\left(\frac{\Lambda_1\Lambda_2}{\Lambda^2_3}\right)^{\frac{1}{3}}ds^2(\text{AdS}_3)+\frac{\rho^{\frac{2}{3}}}{m^2}\bigg[\frac{1}{\gamma^2}\left(\frac{\Lambda_1\Lambda_3}{\Lambda^2_2}\right)^{\frac{1}{3}}ds^2(\text{S}^3)+\frac{1}{(1-\gamma)^2}\left(\frac{\Lambda_2\Lambda_3}{\Lambda^2_1}\right)^{\frac{1}{3}}ds^2(\tilde{\text{S}}^3) \nn \\[2mm]
& + \Lambda_0 (\Lambda_1\Lambda_2\Lambda_3)^{\frac{1}{3}}(d\rho^2+dx^2)\bigg], \nn \\[2mm]
\frac{G}{L^3}&=\bigg(d\left(\frac{\partial_x h}{\Lambda_0 } \frac{1}{\Lambda_3}\right)+2 \rho(\partial_x hd\rho- \partial_{\rho}h dx)\bigg)\wedge \text{vol}(\text{AdS}_3) \label{eq:largeclass} \\[2mm]
+&\frac{1}{m^3}\bigg(d\left(\frac{\partial_x h}{\Lambda_0 } \frac{1}{\gamma^3 \Lambda_2}-\frac{2x}{\gamma^2}\right)-\frac{2 \rho}{\gamma}(\partial_x hd\rho- \partial_{\rho}h dx)\bigg)\wedge \text{vol}(\text{S}^3) \nn \\[2mm]
+&\frac{1}{m^3}\bigg(d\left( \frac{\partial_x h}{\Lambda_0} \frac{1}{(1-\gamma)^3\Lambda_1}+\frac{2x}{(1-\gamma)^2}\right)-\frac{2 \rho}{1-\gamma}(\partial_x hd\rho- \partial_{\rho}h dx)\bigg)\wedge\text{vol}(\tilde{\text{S}}^3), \nn 
\end{align}
where we have introduced 
\begin{align}
\Lambda_0 = (\partial_{x}h)^2 + (\partial_{\rho}h)^2,
\end{align}
to make the expressions more compact and $\Lambda_{1,2,3}$ are the same functions we defined in \eqref{Lambdatoda} that after the coordinate transformation take the form 
\begin{equation}
\begin{split} 
\Lambda_1 = \gamma-\frac{\partial_{\rho} h }{\rho \Lambda_0 }, ~~~ \Lambda_2 = 1-\gamma+\frac{\partial_{\rho} h }{\rho \Lambda_0}, ~~~ \Lambda_3 = \gamma(1-\gamma)-\frac{ 1 + (1 -2 \gamma) \rho \partial_{\rho} h }{\rho^2 \Lambda_0 }. \label{eq:Lambdaslargeclass}
\end{split}
\end{equation}
Solutions in this mildly restricted (with respect to the Toda case) class are now in 1 to 1 correspondence with the solutions of an axially symmetric cylindrical Laplace equation in 3 dimensions
\beq
\frac{1}{\rho}\partial_{\rho}(\rho \partial_{\rho}h)+ \partial_{x}^2h =0.\label{eq:cylindricalLaplace}
\eeq
We note that this equation, but not the metric, is invariant under
\beq
h\to c_1 h+ c_2\log\rho,\label{eq:shits}
\eeq
for $c_{1,2}$ constants. In terms of this new parametrisation the $\alpha \to \frac{1}{\alpha}$ isomorphism of $\mathfrak{d}(2,1;\alpha)$ is realised as
\beq
\gamma\to 1-\gamma,~~~~h\to-h~~~~ L \to - L,
\eeq
which effectively switches S$^3$ and $\tilde{\text{S}}^3$ in \eqref{eq:largeclass}.

\subsection{Recovering two small ${\cal N}=(4,4)$ classes as limits}\label{sec:limitsmall}
We will now comment on a connection between the class of this section and those of sections \ref{eq: T3class} and \ref{eq:wickrotatedclass}. In particular it follows that for every large ${\cal N}=(4,4)$ solution there are 2 corresponding small (4,4) solutions, one with two 3-sphere, one with one of these replaced by $\mathbb{T}^3$.\\
~~\\
We begin with the class of section \ref{eq: T3class}: We note that fixing  $\gamma= 1$ in \eqref{eq:Lambdaslargeclass} makes these expressions reproduce \eqref{eq:Lambdassmallclass}. Of course the warp factor of $\tilde{\text{S}}^3$ blows up in this limit, but it does so in a regular fashion such that  $\tilde{\text{S}}^3\to\mathbb{R}^3$. Thus if we expand \eqref{eq:largeclass} to leading order about $\gamma= 1$ assuming $h$ is regular and identify in this limit
\beq
\frac{1}{(1-\gamma)^2}ds^2(\tilde{\text{S}}^3)= ds^2(\mathbb{R}^3),~~~~\frac{1}{(1-\gamma)^3}\text{vol}(\tilde{\text{S}}^3)= \text{vol}(\mathbb{R}^3),
\eeq
all dependence on $\gamma$ drops out of the class, and $\tilde{\text{S}}^3\to \mathbb{R}^3$.
Of course we are free to compactify $\mathbb{R}^3$ to $\mathbb{T}^3$ as locally they are the same, upon doing this one reproduces precisely  \eqref{eq:smallclassII} and \eqref{eq:Lambdassmallclass}.\\
~~\\
In \cite{Capuozzo:2024onf} it was recently shown that one can realise a class of small ${\cal N}=(4,4)$ solutions as a limit of a class of  large ${\cal N}=(4,4)$ solutions while maintaining both 3-spheres. We can achieve the same in terms of our classification by sending
\beq
h\to -(h+ \frac{1}{1-\gamma}\log\rho),~~~~L\to \frac{L}{(1-\gamma)^{\frac{1}{3}}},
\eeq
and then expanding about $\gamma=1$ assuming $h$ is regular. We find to leading order that the metric and fluxes become
\begin{align}
\label{eq:LimitsmallS3}
\frac{ds^2}{L^2}& =\frac{1}{\left(\rho^{-1}\partial_{\rho}h\right)^{\frac{1}{3}}}\bigg[ds^2(\text{AdS}_3)+\frac{1}{m^2}ds^2(\text{S}^3)\bigg]+\frac{\left(\rho^{-1}\partial_{\rho}h\right)^{\frac{2}{3}}}{m^2}\left(d\rho^2+dx^2+\rho^2 ds^2(\tilde{\text{S}}^3)\right), \\[2mm]
\frac{G}{L^3} & = 2 \, dx \wedge \text{vol}(\text{AdS}_3) - \frac{2}{m^3} \, dx \wedge \text{vol}(\tilde{\text{S}}^3) - \frac{\rho^3}{m^3} \left( \partial_{\rho}(\rho^{-1} \partial_{\rho} h) dx - \partial_{x}(\rho^{-1} \partial_{\rho} h) d\rho \right) \wedge \text{vol}(\text{S}^3). \nn 
\end{align}
We note that $\tilde{h}=\rho^{-1}\partial_{\rho}h$ appears in precisely the same position as $h$ does in \eqref{eq:class3small}. Further when $h$ solves  \eqref{eq:cylindricalLaplace} it follows that $\tilde{h}$  solves
\beq\label{eq:PDELimitsmallS3}
\frac{1}{\rho^3}\partial_{\rho}(\rho^3\partial_{\rho}\tilde{h})+\partial_{x}^2\tilde{h}=0,
\eeq
we thus see that we have reproduced the class of section \ref{eq:wickrotatedclass} with $u=1$ in this limit. However notice that this identification hinges on the fact that a solution to \eqref{eq:cylindricalLaplace} implies a solution to \eqref{eq:PDELimitsmallS3} which negates the need to actually take this limit, which might be awkward for some concrete solutions (specifically the assumption that $h$ is regular at $\gamma=1$).

\subsection{Relation to an existing class of large ${\cal N}=(4,4)$ solutions}\label{sec:otherclass}
In literature there are already some articles that classified solutions with large ${\cal{N}}=(4,4)$ supersymmetry in M-theory \cite{Estes:2012vm,Bachas:2013vza}. In this subsection we will show how our class is compatible with theirs.\\
~\\
We will refer to specifically to \cite{Bachas:2013vza}, which is governed by two functions $G$, $h_{\beta}$ and three constants $\beta_i$. $G$ is a complex function (and should not be confused with the 4-form flux), $h_{\beta}$ is a real harmonic function and both of them have support on the Riemann Surface. The three constants $\beta_1$, $\beta_2$ and $\beta_3$ are defined such that $\beta_1 + \beta_2 + \beta_3 = 0 $. We will call the parameter $\gamma$ and the function $h$ that appear in \cite{Bachas:2013vza}, and conflict with our earlier definitions, $\gamma_{\beta}$ and $h_{\beta}$. To have a supersymmetric solution $(h_{\beta},G)$ must solve following PDEs
\begin{align}
\partial_{\omega} \partial_{\bar{\omega}} h_{\beta} = 0, \qquad \partial_{\omega} G = \frac{1}{2} (G + \bar{G}) \partial_{\omega} \text{ln} \, h_{\beta}.\label{eq:Gpde}
\end{align}
One can embed our large (4,4) class inside that of \cite{Bachas:2013vza} by making the following identifications: The parameters $\beta_i$ and $\gamma_{\beta}$ are related as
\begin{align}
\beta_1 = -1, \qquad \beta_2 = \gamma, \qquad \beta_3 = 1- \gamma, \qquad \gamma_{\beta} = \frac{\beta_2}{\beta_3} = \frac{\gamma}{1 - \gamma}.
\end{align}
The complex coordinates and the functions $h_{\beta}$ and $G$ are
\begin{align}
\omega =-i( x+i \rho), \qquad h_{\beta} = - \frac{L^3}{m^3} \rho, \qquad G = 2 \gamma  \left((1-\gamma ) \rho  \partial_{x} h + i \left( 1 + (1-\gamma ) \rho  \partial_{\rho} h \right)\right)-i,
\end{align}
where $h_{\beta}$ is clearly harmonic and \eqref{eq:Gpde} reduces to \eqref{eq:cylindricalLaplace}. 

Note that \eqref{eq:Gpde} is invariant under $G \to G = i a + b G$, for $(a,b)$ constant, which one can show is equivalent to \eqref{eq:shits}. Using this symmetry we can map $G$ to the simpler form
\begin{align}
G = i \left( 1 - 2 \rho \partial_{\rho} h \right) - 2 \rho \partial_x h.
\end{align}
for $b = - \frac{1}{(1-\gamma)\gamma}$ and $a = -\frac{(1-\gamma ) (1-2 \gamma )-\gamma ^2}{(1-\gamma ) \gamma } $.\\
~~\\
It is worth making some comments about the relative generality of the class of \cite{Bachas:2013vza} and that of section \eqref{eq:laplaceversion}. We have made two assumptions that \cite{Bachas:2013vza} do not. First, when deriving the ${\cal N}=(1,1)$ conditions that we build our ${\cal N}=(4,4)$ classes on top of we have  assumed that AdS$_3$ does not get necessarily enhanced to AdS$_4$. One might worry that this excludes the possibly of finding solutions that are locally AdS$_4\times$S$^7$ in our class -- however the key word in the proceeding sentence is ``necessarily'', indeed we are able to recover such solutions in section \ref{sec:ad4}.
Second the class of section \eqref{eq:laplaceversion} does not contain one true AdS$_3$ solution which \cite{Bachas:2013vza} does, AdS$_3\times\text{S}^3\times \tilde{\text{S}}^3\times \mathbb{T}^2$, although this can be obtained from our earlier Toda system\footnote{Indeed AdS$_3\times\text{S}^3\times \tilde{\text{S}}^3\times \mathbb{T}^2$ is the only solution to the Toda system with $e^D=$ constant. When moving to the Laplace system we explicitly assume that $e^D$ is not constant, which is why the Toda system contains this solution but the Laplace system does not.}. As such \cite{Bachas:2013vza} is the more general classification, but only in so far as it additionally contains one well know symmetric space solution. For all other solutions the classes should be locally equivalent and, at least in our opinion,  a Laplacian is an easier equation to solve than \eqref{eq:Gpde}.

\section{Explicit solutions}\label{sec:examples}
In this section we provide some explicit solutions contained within the various classes we derived in earlier sections in this work.

\subsection{Small ${\cal{N}}=(4,4)$: AdS$_3 \times$ S$^3 \times$ $\mathbb{T}^4 $ foliated over an interval. }\label{sec:smallex1}
We start by deriving  solutions within the class of \eqref{eq:smallclassI} that are a warped product of AdS$_3 \times$ S$^3 \times$ $\mathbb{T}^4\times {\cal I}$: We choose the function $u$ to be linear in $\rho$, $u = \rho$ and $h$ to not depend on $\mathbb{R}^4$, i.e $h=h(\rho)$. We shall additionally compactify $\mathbb{R}^4\to \mathbb{T}^4$ whose isometries are respected. The background becomes
\begin{align}
ds^2 &=  \Delta_1^{\frac{1}{3}} \bigg[\frac{\rho^{\frac{2}{3}}}{h^{\frac{1}{3}}} \left( ds^2(\text{AdS}_3) + \frac{1}{m^2 \Delta_1} ds^2(\text{S}^3) \right)+ \frac{1}{m^2}\frac{h^{\frac{2}{3}}}{\rho^{\frac{4}{3}}} \left(d\rho^2+ \rho \, l^2 \,ds^2(\mathbb{T}^4) \right)\bigg],\nn\\[2mm]
G &=  d \left( \frac{\rho}{2  h}+2 \rho  \right)\wedge \text{vol}(\text{AdS}_3)+ 
 \frac{1}{m^3} d\left( \frac{\rho}{2  h \Delta_1}-2\rho  \right)\wedge \text{vol}(\text{S}^3)+ \frac{l^4}{m^3} h' \text{vol}(\mathbb{T}^4) , \\[2mm]
\Delta_1&=1+\frac{1}{4 h},\nn
\end{align}
where $l$ is a constant we have extracted for convenience and $\text{Vol}(\mathbb{T}^4)=(2\pi)^4$.
Away from the loci of sources the Bianchi identities are implied by
\beq
h''=0,
\eeq
making $h$ locally a linear function.

In what follows we will be interested in the magnetic Page fluxes, i.e (locally) closed fluxes that imply the magnetic Bianchi identities and give well defined charges. The magnetic four form $G_4$ is defined as
\begin{align}
G_4 = G_4 (\text{S}^3) + G_4 (\mathbb{T}^4) = dC_3(\text{S}^3) + G_4 (\mathbb{T}^4),
\end{align}
where 
\begin{align}
C_3(\text{S}^3) = \frac{1}{m^3} \left( \frac{\rho}{2  h \Delta_1}-2(\rho-k)  \right)\wedge \text{vol}(\text{S}^3),\label{eq:CS3}
\end{align}
and is already closed, so can be used to define a charge. $\star_8G_1$ is different, it obeys
\beq
0= d \star_8 G_1 + \frac{1}{2} G_4 \wedge G_4
\eeq
It will be useful to introduce a closed seven form $\hat{G}_7$ in the following way. We have that
\begin{align}
& d \star_8 G_1 + \frac{1}{2} G_4 \wedge G_4 = d \star_8 G_1 + G_4(\text{S}^3) \wedge G_4(\mathbb{T}^4) = d \star_8 G_1 + d C_3 (\text{S}^3) \wedge G_4(\mathbb{T}^4).  \nn
\end{align}
So the following 7-form is closed
\begin{align}
\hat{G}_7 = \star_8 G_1 + C_3 (\text{S}^3) \wedge G_4(\mathbb{T}^4),
\end{align}
Substituting the explicit values of the objects appearing on the RHS of this expression we find
\begin{align}
\hat{G}_7 =  \frac{2 \, l^4 }{m^6} \left(h-(\rho -k) h' \right) \text{vol}(\mathbb{T}^4) \wedge \text{vol}(\text{S}^3).
\end{align}
It then follows that 
\begin{align}
\label{eq:BianchiPage}
d G_4 = \frac{l^4}{m^3} h'' \, d \rho \wedge \text{vol}(\mathbb{T}^4), \qquad d \hat{G}_7 =\frac{2 \, l^4 }{m^6} (\rho -k) h''  \text{vol}(\mathbb{T}^4) \wedge \text{vol}(\text{S}^3) \wedge d\rho,
\end{align}
so away  from sources, for which  $h''=0$, $G_4,\hat G_7$ are indeed closed. But these expressions also have the benefit of being easily interpretable in the presence of sources as we shall see shortly.\\
~\\
\textbf{Boundary behaviour}\\
We would like to construct AdS$_3$ solutions with a bounded internal space, which requires that the interval spanned by $\rho$ terminates at an upper and lower boundary where the solution exhibits either a regular zero or physical singularity. One place to impose such a boundary is at $\rho=0$: As $h$ is a linear function, we can distinguish between two cases depending on whether $h$ does or does not have an order one zero at this loci. If it does $h= h'(0)\rho$ for $h'(0)$ a non zero constant, if it does not $ h= h(0)$ a constant. A boundary can also happen at $\rho = \rho_0>0$ if $h$ has an order one zero at that loci. In this case it takes the form $ h=h'(\rho_0)(\rho-\rho_0)$ for $h'(\rho_0)$ constant. 

We start by studying the limit $\rho \to 0$ in the case that $ h \to h(0)\neq 0$: Upon redefining $\rho = z^2 $ we find that the metric becomes
\begin{align}
ds^2 \sim \frac{z^{4/3}}{h(0)^{1/3}}ds^2(\text{AdS}_3) + 4 \frac{z^{-2/3}}{m^2} h(0)^{2/3} \left( d z^2 + \frac{1}{4 h(0) +1} z^2 ds^2(\text{S}^3) + \frac{l^2}{4} ds^2(\mathbb{T}^4) \right), 
\end{align}
to leading order. This is the behaviour of a partially localised stack of M$2$-brane at conical defect that is extended in  AdS$_3$ and smeared over $\mathbb{T}^4$. We can interpret the factor $\delta = \frac{1}{1+4 h(0)}$ as the deficit angle. When $ \rho \to \rho_0$ such that  $ h\to h'(\rho_0)(\rho-\rho_0)$ we find
\begin{align}
ds^2 =  h'(\rho_0)^{-2/3}(\rho_0 - \rho)^{-2/3} ds^2(\text{AdS}_3)+ \frac{h'(\rho_0)}{m^2 } (\rho_0 - \rho)^{1/3} \left(  d\rho^2 + \rho_0^2 ds^2( \text{S}^3 ) + l^2 \rho_0 ds^2( \mathbb{T}^4) \right),
\end{align}
which is the behaviour of a partially localised O$2$-plane extended in AdS$_3$ and smeared over $\mathbb{T}^4$ .
In the case that $\rho\to 0$ with  $h \to 0 $, we can redefine $\rho = \frac{1}{4} z^2$ and the metric becomes
\begin{align}
ds^2 \sim h'(0)^{-2/3} ds^2 (\text{AdS}_3) + \frac{h'(0)^{1/3}}{m^2} \left( dz^2 + z^2 ds^2(\text{S}^3) +l^2ds^2 (\mathbb{T}^4) \right),
\end{align}
which gives a regular zero. \\
~\\
\textbf{Solution bounded between M2-branes and O2-planes}\\
We can derive a simple bounded solution by taking $h=L (\rho_0 - \rho)$. The interval spanned by $\rho$ is then bounded between an O$2$-plane at $\rho=\rho_0$ and M$2$ at a conical defect at $\rho=0$. We can define a Page charge for M5  and M2 branes by integrating the Page fluxes on $\mathbb{T}^4$ and $(\mathbb{T}^4,\text{S}^3)$ respectively,
\begin{align}
Q_{M5} = \frac{1}{(2 \pi)^3} \int G_4 = \frac{2 \pi  l^4 }{m^3} L , \qquad Q_{M2} = \frac{1}{(2 \pi)^6} \int \hat G_7 = \frac{l^4 }{m^6} L (\rho_0 - k). 
\end{align}
where one can tune the constants such that $Q_{M5}$ and $ Q_{M2}$ are elements of $\mathbb{Z}$. One might also consider integrating $G_4$ on $(\text{S}^3, \rho)$, but the result yields zero, ie
\begin{align}
\tilde{Q}_{M5}= \int_{\rho \times \text{S}^3} G_4 = \int_{\text{S}^3} C_3 \bigg|_{\rho=0}^{\rho=\rho_0} =0.
\end{align}
~\\
\textbf{Global solutions with interior M5 branes}\\
We can consider more general scenario by allowing $h$ to be only piecewise linear such that it is linear for $\rho \in (k,k+1)$, for $k$ an integer, but $h'$ is permitted to be discontinuous at any $\rho=k$. To realise a bounded solution in this manner $h$ must decompose as
\begin{equation}\label{piecewisefunction}
h =
    \begin{cases}   
     ~~~~~~~~~~~~~~~ n_0 + n_1 \rho & \rho \in[0,1]\\
      n_k+(n_{k+1}-n_k)(\rho-k) & \rho  \in[k,k+1]\\
   ~~~~~~~~~   n_{P}(P+1-\rho )  & \rho  \in[P,P+1],
       \end{cases}
\end{equation}
where the coefficients of $h$ are chosen such that $h$ is continuous though its derivative need not be such  the space terminates at a $\rho=P+1$ with an O2 plane and begins at $\rho=0$ with either a regular zero for $n_0=0$ or an M2 brane for $n_0\neq 0$ - note that $n_{-1}=n_{P+1}=0$. We now take \eqref{eq:CS3} to be the 3-form potential on S$^3$ in the $k$'th cell, so that it receives large gauge transformations as one crosses between cells. Given this the Pages charge in the  $k$'th cell are 
\begin{align}
Q_{M2k} = n_k, \qquad Q_{M5k} =  (n_k-n_{k+1} ),  
\end{align}
where we fix $2 \pi  l^2 = 2 \pi  m^3 = 1 $,  flux quantisation then demanding that $n_k$ are integer. We again find
\begin{align}
\tilde{Q}_{M5} = \int_{(\rho ,\text{S}^3)} G_4 = \int_{\text{S}^3} C_3 \bigg|_{\rho=0}^{\rho=P+1} = \sum_{k=0}^{P} \left( \frac{k}{4 n_k +1}+\frac{4 n_{k+1} - k}{4 n_{k+1}+1} \right)= 0.
\end{align}
In general $h'$ can be discontinuous at $\rho=k$, in which case 
\begin{align}
h''_k = m_k \, \delta(\rho - k), \qquad \text{with} \qquad m_k =   n_{k+1} +n_{k-1}-2 n_k.
\end{align}
thus we have a delta function source at $\rho=k$ when $m_k \neq 0$. In terms of which \eqref{eq:BianchiPage} becomes
\begin{align}
d G_4 = \frac{m_k}{2 \pi} \, \delta(\rho - k) \, d \rho \wedge \text{vol}(\mathbb{T}^4),
\end{align}
at this loci. We thus see we have an explicit source for M5 branes of charge $m_k$ while $d \hat{G}_7$ vanishes as $(\rho -k) \delta(\rho - k) \to 0$ meaning there are no source M2 branes. 

\subsection{Small ${\cal{N}}=(4,4)$: AdS$_3 \times$ S$^3 \times$ $\mathbb{T}^5 $ with localised sources}\label{sec:smallex2}
In this subsection we are interested in finding a different solution. We still focus on the class \eqref{eq:smallclassI}, but in this case we will describe the back-reaction of M5 branes and O5 planes on a 5-torus. We follow the same ideas as \cite{Legramandi:2019ulq}. In order to achieve our goal we set the function $u=1$: in this way the PDE \eqref{eq:smallclass1pde} becomes
\begin{equation}\label{eq:T5pde}
\nabla^2_{\mathbb{T}^5} h =0,
\end{equation}
where we compactified the $\mathbb{R}^5$ to be the five dimensional torus $\mathbb{T}^5$. The coordinates have been periodically identified $x_i \cong x_i + 2 \pi $ such that Vol$(\mathbb{T}^5) = (2 \pi)^5$. To describe branes and planes we need to introduce sources $\sigma_j$ in \eqref{eq:T5pde}, the simplest scenario is to introduce only two as
\begin{equation}
\nabla^2_{\mathbb{T}^5} h = \sigma_1 + \sigma_2, \label{eq:T5pdewithsourcessigma}
\end{equation}
where $\sigma_j =(2 \pi)^3 \, N_j \, \delta_{\mathbb{T}^5} (\vec{x}-\vec{x}_j)$ for $j=(1,2)$. The Dirac delta on $\mathbb{T}^5$ can be written as
\begin{equation}
\delta_{\mathbb{T}^5} (\vec{x}-\vec{x}_j) = \sum_{\vec{k} \in \mathbb{Z}^5} \delta( (\vec{x}-\vec{x}_j) - 2 \pi \vec{k} )= \frac{1}{(2 \pi)^5} \sum_{\vec{k} \in \mathbb{Z}^5} \text{exp} \left[ i \, \vec{k} \cdot ( \vec{x}-\vec{x}_j ) \right].
\end{equation}
If we plug this in \eqref{eq:T5pdewithsources} and we integrate on $\mathbb{T}^5$ we get the condition $N_1+ N_2=0$. This is telling us that the right objects are indeed a single M5 brane and O5 plane, since their charges sum to zero. We can rewrite the PDE as
\begin{equation}
\nabla^2_{\mathbb{T}^5} h = (2 \pi)^3  \left(\delta_{\mathbb{T}^5} (\vec{x}-\vec{x}_1) - \delta_{\mathbb{T}^5} (\vec{x}-\vec{x}_2) \right). \label{eq:T5pdewithsources}
\end{equation}
In this way we can derive the explicit solution by simply integrating
\begin{equation}
h = h_0 - \frac{1}{(2 \pi)^2} \sum_{\vec{k} \in \mathbb{Z}^5 - \{\vec{0}\}} \frac{1}{k^2}\left( \text{exp} \left[ i \, \vec{k} \cdot ( \vec{x}-\vec{x}_1 ) \right] - \text{exp} \left[ i \, \vec{k} \cdot ( \vec{x}-\vec{x}_2 ) \right] \right).
\end{equation}
This is actually a bit of an over simplification, as the series defining the above $h$ is not absolutely convergent. A more careful treatment of the general problem of solving Laplace equations on Tori can be found in \cite{Andriot:2019hay,Andriot:2021gwv}, which leads to a solution in terms of Jacobi theta functions.

\subsection{Large ${\cal N}=(4,4)$ solution and asymptotic AdS$_7$}\label{sec:largeex2}
In this section we provide several examples of solutions that preserve Large ${\cal{N}}=(4,4)$ supersymmetry. We make use of a coordinate transformation of the Riemann surface directions that is well suited for recovering AdS$_7$.\\
~\\
We would like to be able to recover the recent solution found in \cite{Conti:2024rwd}. To this end it is helpful to perform the following coordinate transformations 
\beq
x = - \cosh x_1 \sin x_2 , \qquad \rho =  \sinh x_1 \cos x_2,
\eeq
 upon which the governing PDE becomes
\begin{align}\label{eq:transfPDE}
\sinh x_1 \, \partial_{x_2} \left( \cos x_2 \partial_{x_2} h \right) + \cos x_2 \, \partial_{x_1} \left( \sinh x_1 \partial_{x_1} h \right) = 0,
\end{align}
which is invariant under $h\to c_1 h+c_2 \log(\sinh x_1 \cos x_2)$ for $c_{1,2}$ constant. 

We would like to now write the class of section \ref{eq:laplaceversion} in terms of these new coordinates for which the following are useful
\begin{equation}
\begin{split}
\partial_{x} h & = - 2\frac{ \sinh x_1 \sin x_2 \partial_{x_1} h +\cosh x_1 \cos x_2 \partial_{x_2} h }{\cosh 2 x_1 + \cos 2 x_2} ,  \\[2mm]
\partial_{\rho} h & = 2\frac{\cosh x_1 \cos x_2 \partial_{x_1} h-\sinh x_1 \sin x_2 \partial_{x_2} h }{\cosh 2 x_1 + \cos 2 x_2}.
\end{split}
\end{equation}
We find that the metric and flux become
\begin{align}
\frac{ds^2}{L^2}&=(\sinh x_1\cos x_2)^{\frac{2}{3}}\left(\frac{\Lambda_1\Lambda_2}{\Lambda^2_3}\right)^{\frac{1}{3}}ds^2(\text{AdS}_3)+\frac{(\sinh x_1\cos x_2)^{\frac{2}{3}}}{m^2}\bigg[\frac{1}{\gamma^2}\left(\frac{\Lambda_1\Lambda_3}{\Lambda^2_2}\right)^{\frac{1}{3}}ds^2(\text{S}^3) \nn \\[2mm]
&+\frac{1}{(1-\gamma)^2}\left(\frac{\Lambda_2\Lambda_3}{\Lambda^2_1}\right)^{\frac{1}{3}}ds^2(\tilde{\text{S}}^3) + \tilde{\Lambda}_0 (\Lambda_1\Lambda_2\Lambda_3)^{\frac{1}{3}}(dx_1^2+dx_2^2)\bigg], \nn \\[2mm]
\frac{G}{L^3}&=-\bigg(d\left(\frac{\Lambda_4}{\tilde{\Lambda}_0\Lambda_3 } \right)+2 \sinh x_1 \cos x_2(\partial_{x_2} hdx_1- \partial_{x_1}h dx_2)\bigg)\wedge \text{vol}(\text{AdS}_3) \label{eq:largeclass2} \\[2mm]
-&\frac{1}{m^3}\bigg(d\left(\frac{\Lambda_4}{\gamma^3\tilde{\Lambda}_0\Lambda_2 }+\frac{2\cosh x_1\sin x_2}{\gamma^2}\right) - \frac{2}{\gamma}\sinh x_1 \cos x_2(\partial_{x_2} hdx_1- \partial_{x_1}h dx_2)\bigg)\wedge \text{vol}(\text{S}^3) \nn \\[2mm]
-&\frac{1}{m^3}\bigg(d\left(\frac{\Lambda_4}{(1-\gamma)^3\tilde{\Lambda}_0\Lambda_1 }-\frac{2\cosh x_1\sin x_2}{(1-\gamma)^2}\right) - \frac{2 }{1-\gamma}\sinh x_1 \cos x_2(\partial_{x_2} hdx_1- \partial_{x_1}h dx_2)\bigg)\wedge\text{vol}(\tilde{\text{S}}^3), \nn 
\end{align}
where
\begin{equation}
\begin{split}
\Lambda_1&= \gamma-\frac{\coth x_1\partial_{x_1}h-\tan x_2\partial_{x_2}h}{\tilde{\Lambda}_0},~~~\Lambda_2= 1-\Lambda_1,~~~\Lambda_3= \Lambda_1\Lambda_2-\left(\frac{\Lambda_4}{\cosh x_1\sin x_2\tilde \Lambda_0}\right)^2 \\[2mm]
\Lambda_4&= \cos x_2\cosh x_1\partial_{x_2}h+\sinh x_1 \sin x_2\partial_{x_1}h,~~~\tilde{\Lambda}_0= (\partial_{x_1}h)^2+ (\partial_{x_2}h)^2,
\end{split}
\end{equation}
which summaries the content of \eqref{eq:largeclass} in the new coordinates. \\
~\\
We now observe that a simple solution to \eqref{eq:transfPDE} is given by\footnote{One could add a constant term to $h$, but this would have no effect as the metric only depends on its derivatives.}
\begin{align}
\partial_{x_1} h = \frac{1}{\sinh {x_1}}, \qquad \partial_{x_2}h=0,~~~\Rightarrow~~~ h=\log\tanh\left(\frac{x_1}{2}\right),
\end{align}
but given the invariance of the defining PDE mentioned below \eqref{eq:transfPDE} we know that 
\beq \label{eq:defh}
h= c_1 \log\tanh\left(\frac{x_1}{2}\right)+c_2 \log(\sinh x_1 \cos x_2),
\eeq
also solves  \eqref{eq:transfPDE} - this gives a 2 parameter family of solutions to \eqref{eq:transfPDE} which contains several distinct physical solutions.

First it is possible to recover AdS$_7\times$S$^4$ by tuning
\beq
\gamma=\frac{1}{c_1}=\frac{1}{c_2}=2,
\eeq
one then finds that \eqref{eq:largeclass2} reduces to
\begin{align}
\frac{ds^2}{L^2}&=\cosh^2\left(\frac{x_1}{2}\right)ds^2(\text{AdS}_3)+\frac{1}{m^2}\sinh^2\left(\frac{x_1}{2}\right)ds^2(\tilde{\text{S}}^3)+\frac{1}{4m^2}dx_1^2+ \frac{1}{4m^2}\left(dx_2^2+ \cos^2x_2 ds^2(\text{S}^3)\right),\nn\\[2mm]
\frac{G}{L^3}&=-\frac{3}{4m^3}\cos^3 x_2 dx_2\wedge \text{vol}(\text{S}^3),
\end{align}
where we have rescaled $L\to 2^{-\frac{1}{3}}L$, which is of course the claimed geometry with AdS$_7$ expressed as a foliation of AdS$_3\times \tilde{\text{S}}^3$ over an interval.

Second we recover the solution of \cite{Conti:2024rwd} by fixing
\begin{equation}
c_1 = \frac{1}{\gamma \left( \gamma -1\right)}, \qquad c_2 = \frac{1}{1-\gamma}.
\end{equation}
This solution depends on a parameter $\lambda$ and interpolates between a regular zero at $x_1=0$ and AdS$_7\times\text{S}^4$ as $x_2\to \infty$  and  $-\frac{\pi}{2}\leq x_2 \leq \frac{\pi}{2}$. It depends on a parameter $\lambda$ which in terms of our $\gamma$ is
\begin{equation}
\gamma = \frac{\lambda+1}{\lambda-1}.
\end{equation}
That the two solutions really are the same, rather than merely exhibit similar behaviour, is made clear through the identifications
\beq
L^6=\frac{128(1+\lambda)}{g^6(\lambda-1)^6},~~~~x_1=2\text{arctanh}\left(\frac{\cos\theta}{1+\lambda \sin\theta}\right),~~~x_2=\xi,
\eeq
upon which one precisely reproduces (4.4) of \cite{Conti:2024rwd}.

Although \eqref{eq:defh} is more general than \cite{Conti:2024rwd}, we have been unable to identify physical bounded behaviour for other tunings of $(c_1,c_2)$.

\subsubsection{Small $(4,4)$ solution from large $(4,4)$ solution }\label{eq:smallfromlarge}
As pointed out in subsection \ref{sec:limitsmall}, if we have a function $h$ that solves the PDE of the Large class, we automatically obtain two new solution that preserve small supersymmetry. We will now generate a class of AdS$_3 \times$ S$^3 \times \tilde{\text{S}^3} \times \Sigma_2$ solutions in this fashion.\\
~\\
For the case with S$^3$  we start by rewriting \eqref{eq:LimitsmallS3} in $(x_1,x_2)$ coordinates. The useful term involving the function $h$ is
\begin{equation}
\tilde{h} = \rho^{-1}\partial_{\rho}h = 2 \tan {x_2}\frac{\coth {x_1} \cot {x_2} \partial_{x_1}h - \partial_{x_2} h }{\cosh {2 x_1} + \cos {2 x_2}}.
\end{equation}
The metric and the fluxes become
\begin{align}\label{eq:LimitsmallS3x1x2}
\frac{ds^2}{L^2} & =\frac{1}{\tilde{h}^{\frac{1}{3}}}\left(ds^2(\text{AdS}_3)+\frac{1}{m^2}ds^2(\text{S}^3)\right) +\notag \\[2mm] 
& + \frac{\tilde{h}^{\frac{2}{3}}}{m^2} \left( \sinh^2 {x_1} \cos^2{x_2} \, ds^2(\tilde{\text{S}}^3) + \frac{\cosh {2 x_1 } + \cos {2 x_2}}{2} \left(dx_1^2+dx_2^2\right) \right), \notag \\[2mm] 
\frac{G}{L^3} & = - 2 \, d(\cosh {x_1} \sin {x_2} ) \wedge \text{vol}(\text{AdS}_3) + \frac{2}{m^3} \, d (\cosh {x_1} \sin {x_2} ) \wedge \text{vol}(\tilde{\text{S}}^3) \\[2mm]
& - \frac{2}{m^3} \frac{\sinh^4 {x_1} \cos^3 {x_2} \sin {x_2}}{\cosh {2 x_1} + \cos { 2 x_2}} \Big( (\coth {x_1} \cot {x_2} \partial_{x_1}\tilde{h} -\partial_{x_2} \tilde{h}) \, d(\cosh {x_1} \sin {x_2})  \notag \\[2mm]
& + ( \partial_{x_1} \tilde{h} + \coth x_1 \cot x_2 \, \partial_{x_2} \tilde{h} ) \, d (\sinh {x_1} \cos {x_2}) \Big) \wedge \text{vol}(\text{S}^3). \notag
\end{align}
The function $h$ defined in \eqref{eq:defh} implies that a solution of \eqref{eq:LimitsmallS3x1x2} is 
\begin{equation}
\tilde{h} = 2 \frac{c_1 \coth {x_1} \text{csch}{x_1} + c_2 \coth ^2{x_1} + c_2 \tan ^2{x_2} }{\cosh {2 x_1} + \cos {2 x_2 }}.
\end{equation}
Indeed it satisfies \eqref{eq:PDELimitsmallS3} in $(x_1,x_2)$ coordinates
\begin{equation}
\partial_{x_1}^2 \tilde{h}+ \partial_{x_2}^2 \tilde{h} -3 \tan {x_2} \partial_{x_2} \tilde{h} + 3 \coth {x_1} \partial_{x_1}\tilde{h} = 0.
\end{equation}
According to how we tune the constant we can find different global behaviour.

We reproduce the solution found in section 4.2 of \cite{Conti:2024rwd} if we set $c_1 \neq 0$ and $c_2 = 0$. At $x_1 \to 0$ we have M5 brane smeared along $x_2$, while at $x_1 \to \infty$ there is AdS$_7 \times$ S$^4$.

For any other different tuning of the constant we have not been able to make a physical interpretation of the solution.

\subsection{Large ${\cal N}=(4,4)$ solution  and asymptotic AdS$_4$}\label{sec:ad4}
In this section we provide further examples of solutions that preserve Large ${\cal{N}}=(4,4)$ supersymmetry, this time making use of a coordinate transformation of the Riemann surface that is well suited for recovering AdS$_4$. We will in fact find a 3 parameter (in addition to $\gamma$) family of solutions that for certain tunings give rise to both local AdS$_4\times$S$^7$ and also to a class of Janus solutions that tend to this at certain loci derived in \cite{Bobev:2013yra}.\\
~\\
This time we choose to reparametrise 
\beq
\rho=\cosh(2x_1)\sin(2x_2),~~~x=\sinh(2x_1)\cos(2x_2),
\eeq
such that the class is mapped to
\begin{align}
\frac{ds^2}{L^2}&=(\cosh(2x_1)\sin(2x_2))^{\frac{2}{3}}\bigg[\left(\frac{\Lambda_1 \Lambda_2}{\Lambda_3^2}\right)^{\frac{1}{3}}ds^2(\text{AdS}_3)+ \frac{1}{m^2}\bigg(\frac{1}{\gamma^2}\left(\frac{\Lambda_1\Lambda_3}{\Lambda_2^2}\right)^{\frac{1}{3}}ds^2(\text{S}^3)\nn\\[2mm]
&+ \frac{1}{(1-\gamma)^2}\left(\frac{\Lambda_2\Lambda_3}{\Lambda_1^2}\right)^{\frac{1}{3}} ds^2(\tilde{\text{S}}^3)+ \tilde{\Lambda}_0(\Lambda_1\Lambda_2\Lambda_3)^{\frac{1}{3}}(dx_1^2+dx_2^2)\bigg)\bigg],\nn\\[2mm]
\frac{G}{L^3}&=-\bigg(d\left(\frac{\Lambda_4}{\Lambda_3 \tilde{\Lambda}_0}\right)+2 \cosh(2x_1)\sin(2x_2)\left(\partial_{x_2}h dx_1-\partial_{x_1}h dx_2\right)\bigg)\wedge \text{vol}(\text{AdS}_3),\nn\\[2mm]
&-\frac{1}{m^3\gamma}\bigg(d\left(\frac{\Lambda_4}{\gamma^2\Lambda_2\tilde{\Lambda}_0}+\frac{2}{\gamma}\sinh(2x_1)\cos(2x_2)\right)-\frac{2}{L}\cosh(2x_1)\sin(2x_2)\left(\partial_{x_2}h dx_1-\partial_{x_1}h dx_2\right)\bigg)\wedge\text{vol}(\text{S}^3)\nn\\[2mm]
&-\frac{1}{m^3 (1-\gamma)}\bigg(d\left(\frac{\Lambda_4}{(1-\gamma)^2\Lambda_1\tilde{\Lambda}_0}-\frac{2}{(1-\gamma)}\sinh(2x_1)\cos(2x_2)\right)\nn\\[2mm]
&-2\cosh(2x_1)\sin(2x_2)\left(\partial_{x_2}hdx_1-\partial_{x_1}hdx_2\right)\bigg)\wedge\text{vol}(\tilde{\text{S}}^3)\label{eq:AdS4class}
\end{align}
where now
\begin{align}
\Lambda_1&=\gamma-2\frac{\cot(2x_2)\partial_{x_2}h+\tanh(2x_1)\partial_{x_1}h}{\tilde{\Lambda}_0}~~~~\Lambda_2=1-\Lambda_1,~~~~\Lambda_3=\Lambda_1\Lambda_2-\left(\frac{\Lambda_4}{\cosh(2x_1)\sin(2x_2)\tilde{\Lambda}_0}\right)^2,\nn\\[2mm]
\Lambda_4&=2(\sinh(2x_1)\sin(2x_2)\partial_{x_2}h-\cosh(2x_1)\cos(2x_2)\partial_{x_1}h),~~~\tilde{\Lambda}_0=(\partial_{x_1}h)^2+(\partial_{x_2}h)^2,
\end{align}
and the defining PDE gets mapped to
\beq
\sin(2x_2)\partial_{x_1}\left(\cosh(2x_1) \partial_{x_1}h\right)+\cosh(2x_1)\partial_{x_2}\left(\sin(2x_1)\partial_{x_2}h\right)=0.\label{eq:AdS4PDE}
\eeq
It is not hard to verify that 
\beq
h=\frac{c_1}{2} \log(\cosh(2x_1)\cos^2 x_2)+c_2 \log(\cosh(2x_1)\sin(2x_2))+c_3 \text{arctan}(\tanh(x_1)),
\eeq
for $(c_1,c_2,c_3)$ arbitrary constants furnishes us with a solution to \eqref{eq:AdS4PDE}. This allows for different boundary behaviours depending on how $(c_1,c_2,c_3,\gamma)$ are tuned:

First of all one recovers AdS$_4\times $S$^7$ locally by fixing
\beq
\frac{c_1}{2}=-c_2=\frac{1}{\gamma}=2,~~~~c_3=0.
\eeq
upon which \eqref{eq:AdS4class} reduces to 
\begin{align}
\frac{ds^2}{L^2}&=\cosh^2(2x_1)ds^2(\text{AdS}_3)+\frac{dx_1^2}{4m^2}+\frac{4}{m^2}\bigg(dx_2^2+ \cos^2x_2ds^2(\text{S}^3)+ \sin^2x_2ds^2(\tilde{\text{S}}^3)\bigg),\nn\\[2mm]
\frac{G}{L^3}&=6\cosh^3(2x_1)dx_1\wedge \text{vol}(\text{AdS}_3)\label{eq:AdS4ss}
\end{align}
where we have rescaled $L\to  2^{-\frac{1}{3}}L$.  Here the first 2 terms in the metric form $ds^2(\text{AdS}_4)$ with inverse radius $m$, while $G\propto \text{vol}(\text{AdS}_4)$ - the remaining directions form a 7-sphere in topological joint coordinates, yielding the claimed result. 

We find a more interesting class of solutions by tuning
\beq
c_1=\frac{1}{(1-\gamma)\gamma},~~~~c_2=-\frac{1}{1-\gamma},
\eeq
which can be derived by demanding that the AdS$_3$ warp factor does not experience a zero when $\sin(2x_2)=0$. It is possible to show that this solution is regular and tends to AdS$_4\times $S$^7$ as $x_1\to\pm \infty$ - as such it defines a Janus solution dual to an interface. This solution has in fact been found before in \cite{Bobev:2013yra}, so we will not present its rather complicated form explicitly. However one can map this solution to the $d=11$ uplift of the regular solution in section 5 of \cite{Bobev:2013yra} by identifying 
\begin{align}
(x_1,x_2)&=(\frac{\mu}{2},\theta),~~~~\gamma=\frac{1+a\cos\zeta_0}{2},\nn\\[2mm]
c_3&=\frac{4 a \sin\zeta_0}{\sqrt{1-a^2}(1-a^2 \cos^2\zeta_0)},~~~~L^3=\frac{\sqrt{1-a^2}(1-a^2 \cos^2\zeta_0)}{2}.
\end{align}
It would be interesting to explore whether more AdS$_4$ Janus solutions can be constructed in this class and whether there is a sensible solution\footnote{A solution is guaranteed, one with a physical interpretation is not} with small ${\cal N}=(4,4)$ that follow from  \cite{Bobev:2013yra}  in the fashion of section \ref{eq:smallfromlarge}. We leave these possibilities on the table to be explored later.

\section{Conclusions}\label{sec: conclusions}
In this work we have classified warped AdS$_3$ solutions of $d=11$ supergravity that preserve ${\cal N}=(4,4)$ supersymmetry. We achieved this by a two step process
\begin{enumerate}
\item We derived necessary and sufficient differential and algebraic constraints for ${\cal N}=(1,1)$ supersymmetry in terms of an SU(3)-structure their internal space supports.
\item We constructed general SU(3)-structure forms that transform under the R-symmetry of ${\cal N}=(4,4)$. With respect to these (4,4) charged form we then used the ${\cal N}=(1,1)$ supersymmetry conditions to construct the local form of all solutions they are compatible with up to simple PDEs.
\end{enumerate} 
There are two distinct ways in which solutions can realise (4,4) supersymmetry, in terms of two copies of either the large or small ${\cal N}=4$ superconformal algebras. We find a single class supporting the large (4,4) algebra in section \ref{largeclass} and 3 classes supporting its small cousin in sections \ref{eq: R4class}, \ref{eq: T3class} and \ref{eq:wickrotatedclass}
, though the class of section \ref{eq:wickrotatedclass}
, which recovers a  class from \cite{Dibitetto:2020bsh}, is actually contained in that of section \ref{eq: R4class}. 

The small ${\cal N}=(4,4)$ class of section \ref{eq: R4class} is locally a warped product of $\text{AdS}_3\times \text{S}^3\times \mathbb{R}^4\times \mathbb{R}$ depending on two functions, $u$ which is a linear function of $\mathbb{R}$ and $h$ which is a function of $\mathbb{R}^4\times \mathbb{R}$ which obeys a $u$ dependent deformation of the Laplacian on $\mathbb{R}^5$. The class of section \ref{eq:wickrotatedclass} is obtained from this by parametrising the $\mathbb{R}^4$ factor as $(r,\tilde{\text{S}}^3)$ then demanding that $h$ respects the SO(4) isometry of  $\tilde{\text{S}}^3$. We also show that the entire class can be realised as a near horizon limit involving dyonic M5 branes by generalising an earlier construction of \cite{Dibitetto:2020bsh}.

The small ${\cal N}=(4,4)$ class of section \ref{eq: T3class} is a foliation of $\text{AdS}_3\times \text{S}^3\times \mathbb{T}^3$ over a Riemann surface $\Sigma_2$ with warp factors depending on a single function define on $\Sigma_2$ and its derivatives. Solutions in this class are in 1 to 1 correspondence with solutions to the Laplace equation on a 3 dimensional axially symmetric cylindrical, similar to the ${\cal N}=2\neq$ AdS$_5$ class of \cite{Gaiotto:2009gz}.

The large ${\cal N}=(4,4)$ class of section \ref{largeclass} is another foliation over a Riemann surface $\Sigma_2$, this times with  $\text{AdS}_3\times \text{S}^3\times \tilde{\text{S}}^3$ leaves. In general solutions are defined in terms of a function $D$ with support on $\Sigma_2$ that must solve a 2-dimensional Toda equation. The class also depends on a continuous parameter $\gamma$ that we argue realises the constant $\alpha$ in $\mathfrak{d}(2|1;\alpha)$ as  $\alpha=\frac{\gamma}{1-\gamma}$ . If $D$ is constant the only solution is locally $\text{AdS}_3\times \text{S}^3\times \tilde{\text{S}}^3\times \mathbb{T}^2$, while if $D\neq$ constant we are able to perform a coordinate transformation which maps the Toda to the same axially symmetric cylindrical Laplace equation that defines the class of section \ref{eq: T3class}. This results in the marginally less general class of section \ref{eq:laplaceversion}, which omits only the solution $\text{AdS}_3\times \text{S}^3\times \tilde{\text{S}}^3\times \mathbb{T}^2$, but has the benefit of being defined in terms of a more amenable PDE\footnote{Toda equations, while integrable,  are famously difficult to actually find solutions to.}. The class of section \ref{largeclass} provides an alternative parameterisation of solutions preserving large (4,4) to that appearing in the earlier works \cite{Estes:2012vm,Bachas:2013vza}, and we explicitly show how 
our Laplace equation arises from the conditions of \cite{Bachas:2013vza}. Additionally, motived by the findings of \cite{Capuozzo:2024onf}, we establish that it is possible to recover the small (4,4) classes of sections \ref{eq: T3class} and \ref{eq:wickrotatedclass} (with $u=1$) as certain slightly singular limits, involving $\gamma$, of the large (4,4) class defined in terms of a Laplace equation. Our results make clear that the reason this map between large and small classes  works is that they are actually defined by the same PDE.\\
~\\
We have also constructed several explicit solutions that lie within the classes we construct:
\begin{itemize}
\item In section \ref{sec:smallex1} we construct small ${\cal N}=(4,4)$ solutions within the class of section \ref{eq: R4class} that are warped products of $\text{AdS}_3\times \times \text{S}^3\times \mathbb{T}^4\times I$. The interval $I$ can be bounded between either a regular zero or stack of M2 branes at one end and O2 plane at the other. An arbitrary number of smeared M5 branes can also be placed along the interior of the interval with the Page charges for M2 and M5 branes changing between them. 
\item We construct a solution with fully localised M5 and O5 planes extended on AdS$_3\times$S$^3$ and backreacted on $\mathbb{T}^5$ in section \ref{sec:smallex2}, which is another example in the class of of section \ref{eq: R4class}.
\item In section \ref{sec:largeex2} we construct  a 2 parameter (in addition to $\gamma$) family of solutions preserving large ${\cal N}=(4,4)$ supersymmetry.
\item We find a distinct large ${\cal N}=(4,4)$ solution in section \ref{sec:ad4} depending on 3 parameters. 
\end{itemize}
Amongst these, the small ${\cal N}=(4,4)$ solutions of section \ref{sec:smallex1} are particularly interesting as, being bounded, they should have a well defined ${\cal N}=(4,4)$ supersymmetric CFT$_2$ dual. In general these solutions contain smeared source M5 branes localised along the interval with additional M2 and M5 branes suspended between them, as indicated by the Page charges between the source M5 branes. This describes a Hanany-Witten type brane intersections suggesting that the CFT duals should be certain linear quivers. In \cite{Lozano:2019jza,Lozano:2019emq,Lozano:2019zvg,Lozano:2019ywa} similar solutions in massive IIA, albeit preserving only small ${\cal N}=(4,0)$, were constructed where the dual quivers where identified. It would be interesting to pursue such a matching  for these $d=11$ solutions. Additionally, given that these solutions contain an AdS$_3\times$S$^3$ factor, which realises the near horizon limit of an extremal black string in 6 dimensions, they could be fruitful for the counting of microstates in the vien of \cite{Strominger:1996sh}.

One weakness of the solutions in section \ref{sec:smallex1} is that they contains O2 planes that are smeared over the $\mathbb{T}^4$. Smeared O2 planes are not well defined objects in string theory, but we suspect that this problem can be resolved by generalising the solutions such that their warp factor also depends on the $\mathbb{T}^4$ directions. The solution of section \ref{sec:smallex2} is a partial move in this direction where we show that small ${\cal N}=(4,4)$ solutions with fully localised O5 planes are possible, but ideally one would like solutions that maintain the Hanany-Witten intersections while also making the O-planes localised, this could be fruitful to pursue.

It was proven in \cite{Bachas:2013vza} that the only large ${\cal N}=(4,4)$ solutions with bounded internal space are locally $\text{AdS}_3\times \text{S}^3\times \tilde{\text{S}}^3\times \mathbb{T}^2$, as such the main utility of the class in section \ref{eq:laplaceversion}  is in constructing holographic duals to defects and interfaces in higher dimensional CFT's. It is in this realm that we focused on constructing solutions. Specifically the family of solutions in section \ref{sec:largeex2} generalises solutions constructed in \cite{Conti:2024rwd,Conti:2024qgx} that asymptote to AdS$_7$ where they are argued to be dual to defects in CFT$_6$. We were also able to derive a small ${\cal N}=(4,4)$ solution asymptotic to AdS$_7$, also found in \cite{Conti:2024rwd,Conti:2024qgx}, that can be realised as a limit of this large (4,4) family of solutions. The family of solutions in section \ref{sec:ad4} on the other hand generalises a Janus solution that asymptotes to AdS$_4$ and was found in \cite{Bobev:2013yra}. This is dual to an interface between CFT's in 3 dimensions. Although we do achieve a generalisation in each of these solutions they do not appear to lead to anything physical, other than facilitating the limit that realises the asymptotically AdS$_7$ small ${\cal N}=(4,4)$ solution. 

This really only scratches the surface of the possibilities for finding defects and interfaces within our large (4,4) construction. In the future it would be interesting to properly develop the electro statics analogue to these solutions that their defining cylindrical Laplacian suggests. We expect that it should be possible to construct many more solutions through these means. It would be particularly interesting to construct more interfaces in CFT$_3$ similar to \cite{Bobev:2013yra}. It would also be interesting to establish if \cite{Bobev:2013yra} has a small ${\cal N}=(4,4)$ analogue realisable as a limit.

Finally some additional interesting future directions are the following:
\begin{itemize}
\item Completing the classification of  ${\cal N}=(4,4)$ supersymmetric AdS$_3$ by extending our results to type II supergravity. The necessary foundation, i.e conditions for ${\cal N}=(1,1)$ supersymmetric AdS$_3$ was already derived in \cite{Macpherson:2021lbr}.
\item It would be interesting to employ similar methods to those of this work to peruse the status of large ${\cal N}=(4,0)$ solutions in ten and eleven dimensions as not many explicit examples are known.
\item Related to the previous, the large ${\cal N}=(4,4)$ class contains an $\text{AdS}_3\times \text{S}^3\times \tilde{\text{S}}^3$ factor and can thus be reduced to type IIA along the Hopf-fiber of any of these spaces. If one reduces on AdS$_3$ one arrives at a class of large ${\cal N}=4$ AdS$_2$ solutions, if instead one reduces on one of the 3-sphere factors a class of ${\cal N}=(4,0)$ AdS$_3$ is produced. One can also reduce on the sum of the two 3-sphere Hopf fiber directions and arrive at a class of AdS$_3$ preserving ${\cal N}=(4,2)$ supersymmetry on a foliation of $\mathbb{T}^{(1,1)}$ over a Riemann surface. It is reasonable to expect all of these to admit generalisations with non trivial Romans mass and finding them would be very interesting.
\end{itemize}

\section*{Acknowledgements}
We thank Nikolay Bobev for useful comments. AC would like to thank Giuseppe Dibitetto, Yolanda Lozano, Nicol\`o  Petri and Anayeli Ram\'irez for work on related topics that was informative. The work of NM and AC is supported by grants from the Spanish government MCIU-22-PID2021-123021NB-I00 and principality of Asturias SV-PA-21-AYUD/2021/52177. NM is also supported by the Ram\'on y Cajal fellowship RYC2021-033794-I. The work of AC is funded by the Severo Ochoa fellowship PA-23-BP22-019.

\appendix

\section{Conventions}\label{Conventions}
The bosonic sector of $d=11$ supergravity consists of a metric $g$ and a 3-form potential $C$ with field strength $G = dC$. The action is
\begin{equation}
\label{eq:action11D}
S= \frac{1}{2 k^2 } \int R \star_{11} 1 - \frac{1}{2} G \wedge \star_{11} G - \frac{1}{6}G \wedge G \wedge C.
\end{equation}
Our convention for the hodge dual is
\beq
\star e^{\underline{M}_1...\underline{M}_k}=\frac{1}{(d-k)!}\epsilon_{\underline{M}_{k+1}...\underline{M}_{d-k}}^{~~~~~~~~~~~~~\underline{M}_1...\underline{M}_k}e^{\underline{M}_{k+1}...\underline{M}_{d-k}},
\eeq
where the $d=11$ indices $M$ are curved  and $\underline{M}$ are flat. 
The field equations that follow from \eqref{eq:action11D} are
\begin{subequations}
\label{eq:gen11D}
\begin{align}
& R_{M N} - \frac{1}{12} \left( G_{M A B C} \tensor{G}{_N ^A ^B ^C} - \frac{1}{12} g_{M N} G_{ABCD} G^{ABCD} \right) = 0, \\[2mm]
& d \star_{11} G + \frac{1}{2} G \wedge G = 0,
\end{align}
\end{subequations}
and the Bianchi identity is
\begin{align}
\label{Bianchi}
d G = 0.
\end{align}
The Killing spinor equation is
\begin{align}
\nabla_{M} \epsilon + \frac{1}{288} \left( \tensor{\Gamma}{_M ^{ABCD}} - 8 \delta_M ^A \Gamma^{BCD} \right) G_{ABCD} \epsilon = 0.\label{eq:deq10KSE}
\end{align}
The reader may be more familiar to the expression
\begin{align}
\nabla_{M} \epsilon + \frac{1}{24} \left( 3 \slashed{G} \Gamma_M - \Gamma_M \slashed{G} \right) \epsilon = 0,
\end{align}
which is equivalent.

\section{AdS$_3$ spinors and bi-linears}
\label{AdS$_3$ spinors and bi-linears}
In this appendix we review some properties about the AdS$_3$ Killing spinors and the bilinears we can construct from them, such a computation was already carried out in \cite{Macpherson:2021lbr}.
SO(2,2) is the symmetry group of AdS$_3$ and it can be decomposed as SO(2,2) = SL(2)$_+ \times$ SL(2)$_-$, the Killing spinors $\zeta^{\pm}$ which are charge under these two subgroubs SL(2)$_{\pm}$ obey
\begin{align}
\label{KillingAdSappendix}
\nabla_{\mu}\zeta^{\pm}=\pm \frac{m}{2}\gamma_{\mu} \zeta^{\pm}.
\end{align}
We can choose a parametrization where the gamma matrices are real, in this way also the Killing spinors are. From the Killing spinors we can construct the following bilinears on AdS$_3$
\begin{align}
& \bar{\zeta^-} \gamma_a \zeta^- e^a = v^+, \nn \\
& \bar{\zeta^+} \gamma_a \zeta^+ e^a = v^-, \\
& \frac{1}{2} \left( \bar{\zeta^+} \gamma_a \zeta^- e^a + \bar{\zeta^-} \gamma_a \zeta^+ e^a \right) = - u. \nn
\end{align}
In our convention
\begin{align}
\bar{\zeta}^{\pm} = (\gamma_0 \zeta^{\pm})^{\dagger}.
\end{align}
$f$ is a function, while $u$ and $v^{\pm}$ are 1-forms. They obey the following relations
\begin{align}
\label{external derivatives and inner product}
dv^{\pm}&=2 m f^{-1}v^{\pm}\wedge u,& df&=-m u,\nn\\[2mm]
u\lrcorner\text{vol}(\text{AdS}_3)&=\frac{1}{2}f^{-1}v^+\wedge v^-,& v^{\pm}\lrcorner\text{vol}(\text{AdS}_3)&=\pm f^{-1}v^{\pm}\wedge u.
\end{align}
These equations are also connected to the computation of the hodge dual, we report them here since they supplement the future computation
\begin{align}
\label{hodgedual}
& \star_3 v^{\pm} = \pm f^{-1} v^{\pm} \wedge u, \qquad \star_3 u = \frac{1}{2} f^{-1} \left(v^+ \wedge v^- \right), \qquad  \star_3 (v^+ + v^-) \wedge u = - f (v^+ - v^-).
\end{align}
We would like $u$ and $v^{\pm}$ to give rise to Killing vectors or conformal Killing vectors on AdS$_3$, indeed we can check that
\begin{align}
\nabla_{(\mu}v^{\pm}_{\nu)}=0,\qquad \nabla_{(\mu}u_{\nu)} = \frac{1}{2} \left( \nabla_{\mu} u_{\nu} + \nabla_{\nu} u_{\mu} \right) = - m f g(\text{AdS}_3)_{\mu\nu},
\end{align}
proving $(v^{\pm})^{\mu}\partial_{\mu}$ are both Killing vectors while $(u)^{\mu}\partial_{\mu}$ is a conformal Killing vector. We report below the interior product relations of the vectors
\beq
v^{\pm}\lrcorner v^{\pm}=v^{\pm}\lrcorner u=0,\qquad v^{\pm}\lrcorner  v^{\mp}=-2u\lrcorner u =-2f^2,
\eeq
A particular parameterisation of AdS$_3$ is
\beq
\label{AdS3 parameterisation}
ds^2(\text{AdS}_3)= e^{2m r}(-dt^2+ dx^2)+dr^2.
\eeq
Using this parameterisation , the Killing spinors that are Poincar\'e invariant in $(t,x)$ are
\beq
\zeta_+= e^{\frac{m}{2}r}\left(\begin{array}{c} 1\\0\end{array}\right),\qquad \zeta_-=e^{\frac{m}{2}r}\left(\begin{array}{c} 0\\1\end{array}\right),
\eeq
if we choose the three dimensional gamma matrices to be $\gamma_{\mu}=(i \sigma_2,\sigma_1,\sigma_3)_{\mu}$. The 1-forms and $f$ then take the following expression
\beq
f= e^{m r},\qquad v^{\pm}=e^{2m r}(dt\pm dx),\qquad u=- e^{mr} dr.
\eeq

\section{ $\mathcal{N} = (1,1)$ in $d=11$}
\label{d=11 N=(1,1)}
In this appendix we present the necessary and sufficient conditions for an AdS$_3$ vacuum solution in M-theory to preserve $\mathcal{N} = (1,1)$ supersymmetry.\\
~\\
The most general decomposition in M-theory that gives rise to an AdS$_3$ vacuum solution admits the following ansatz
\begin{align}
ds^2 = e^{2 A} ds^2 (\text{AdS}_3) + ds^2 (\text{M}_8), \qquad G = e^{3A} \text{vol} (\text{AdS}_3) \wedge G_1 + G_4
\end{align}
As carried out in \cite{Gauntlett:2002fz}, the necessary conditions for supersymmetry are defined in terms of the following bilinears
\begin{align}
K_M = \bar{\epsilon} \Gamma_M \epsilon, \qquad \Xi_{MN} = \bar{\epsilon} \Gamma_{MN} \epsilon, \qquad \Sigma_{MNOPQ} = \bar{\epsilon} \Gamma_{MNOPQ} \epsilon,\label{eq:10dforms}
\end{align}
where the conventions are such that $\bar{\epsilon} = \epsilon^{\dagger} \Gamma_0$ and $\Gamma_0...\Gamma_{10}=1$. Indeed, it has been shown that the forms must satisfy the following relations in order to preserve supersymmetry
\begin{subequations}
\label{fluxconstraints}
\begin{align}
\nabla_{(M} K_{N) } & = 0, \label{Killing vector} \\[2mm]
\star_{11} dK & =  \frac{2}{3}  \Xi \wedge \star_{11} G  - \frac{1}{3} \Sigma \wedge G  \label{fluxconstraints1} \\[2mm]
d \Xi & = K \lrcorner G, \label{fluxconstraints2} \\[2mm]
d \Sigma & = K \lrcorner \star_{11} G - \Xi \wedge G. \label{fluxconstraints3}
\end{align}
\end{subequations}
In general these conditions are necessary, in the case that $K$ is time-like they become also sufficient. \\
In 11 dimensions we split the gamma matrices in the following way
\begin{align}
\Gamma_{\underline{\mu}} = \gamma_{\underline{\mu}} \otimes \hat{\gamma}, \qquad  \Gamma_{\underline{a}} = \mathbbm{1} \otimes \gamma_{\underline{a}}, \qquad B_{11} = \mathbbm{1}_2 \otimes B_8
\end{align}
where $\gamma_{\mu}$ are the three dimensional gamma matrices defined in appendix \ref{AdS$_3$ spinors and bi-linears} and $\gamma_a$ are a representation of gamma matrices on M$_8$. We used the underline to remember we are working in the vielbein frame. In our convention the chiral gamma $\hat{\gamma}$ is defined as
\beq
\hat{\gamma} =  \gamma_{12345678}.
\eeq
The intertwiner $B_8$, defining Majorana conjugation on M$_8$ as $\chi^c:= B_8 \chi^*$, is such that $B_8^{-1} \gamma_a B= \gamma_a^{\star}$ and $B_8 B_8^*=B_8 B_8 ^{\dagger}= 1$. \\
Finally, the correct ansatz for our Majorana spinor is
\begin{align}
\label{11dspinor}
\epsilon = \zeta^+ \otimes \chi^+ + \zeta^- \otimes \chi^- ,
\end{align}
where $\zeta^{\pm}$ are real Killing spinors on AdS$_3$ obeying \eqref{KillingAdSappendix}, $\chi^{\pm}$ are non chiral Majorana spinors on M$_8$. \\
The 11d KSE get reduced to the d=8 dimensional constraints
\begin{subequations}
\begin{align}
\label{8dKSE}
& \nabla_a \chi^{\pm} + \frac{1}{24} \left( - 3 \hat{\gamma} G_1 \gamma_a + 3 G_4 \gamma_a +\gamma_a \hat{\gamma} G_1 -\gamma_a G_4 \right) \chi^{\pm}=0, \\[2mm]
& \left(  \left( \pm e^{-A} m + \hat{\gamma} dA \right) + \frac{1}{6} \left( 2 G_1 + G_4 \hat{\gamma} \right) \right) \chi^{\pm} = 0.
\end{align}
\end{subequations}
Using \eqref{11dspinor}, it turns out the 1-form can be expressed as
\begin{align}
\label{firstdefinitionofK}
K = -\left( e^A \left( \chi^{+ \dagger} \chi^{+} v^- + \chi^{- \dagger} \chi^{-} v^+ - \left( \chi^{-\dagger} \chi^{+}  + \chi^{+ \dagger} \chi^{-} \right) u \right)+  f \xi \right),
\end{align}
where
\begin{align}
\label{definitionxi}
\xi = 2  \chi^{+ \dagger} \gamma_a \hat{\gamma} \chi^{-} e^a =  \chi^{+ \dagger} \gamma_a \hat{\gamma} \chi^{-} e^a -  \chi^{- \dagger} \gamma_a \hat{\gamma} \chi^{+} e^a .
\end{align}
So we start to study these conditions. We split the indices of \eqref{Killing vector} along the two different spaces. The Killing vector equation on $K$ reduces to the four following equations
\begin{subequations}
\begin{align}
& \nabla_{(a} \xi_{b)} = 0, \label{one}\\[2mm]
& \mathcal{L}_{\xi} A + m e^{-A} (\chi^{- \dagger} \chi^+ + \chi^{+ \dagger} \chi^-) = 0, \label{two} \\[2mm]
& d \left( e^{-A} \left( \chi^{- \dagger} \chi^+ + \chi^{+ \dagger} \chi^-\right) \right) + m e^{-2A}\xi = 0, \label{three}\\[2mm]
& d \left( e^{-A} \left( \chi^{+ \dagger} \chi^+ \right)\right) = d \left( e^{-A} \left( \chi^{- \dagger} \chi^- \right)\right) = 0. \label{four}
\end{align}
\end{subequations}
Using a very similar method, carried out already in \cite{Macpherson:2021lbr} we solve the conditions as
\begin{align}
e^{-A} \left( \chi^{- \dagger} \chi^+ + \chi^{\dagger +} \chi^- \right) = - \rho(r), \qquad m \xi = e^{2 A} \rho' dr,
\end{align}
where $\rho'$ parametrize the diffeomorphism invariance, we fix it with
\begin{align}
\xi^a \partial_a = \partial_r \qquad \Rightarrow \qquad ||\xi||^2 = \frac{1}{m} e^{2 A} \rho'.
\end{align}
This implies
\begin{align}
ds^2(\text{M}^8) = (e^{\xi})^2 + ds^2(\text{M}^7), \qquad e^{\xi} = \frac{\xi}{||\xi||} = e^{A} \sqrt{\frac{\rho'}{m}} dr.
\end{align}
$\xi$ being a Killing vector implies
\begin{align}
e^A = \sqrt{\frac{m}{\rho'}} e^{A_4}, \qquad \partial_r A_4 =0.
\end{align}
We plug this into \eqref{three} and the solution for $\rho$ is
\begin{align}
\rho = \frac{1}{m} \tanh {r},
\end{align}
the 11d metric becomes
\begin{equation}
\begin{split}
ds^2 & = e^{2 A} ds^2(\text{AdS}_3) + e^{2 A} \frac{\rho'}{m} dr^2 + ds^2(\text{M}^7) \\[2mm]
& = e^{2 A_4} \left( m^2 \cosh^2 r \, ds^2(\text{AdS}_3) + dr^2 \right) + ds^2(\text{M}^7),\label{eq:noads3ex}.
\end{split}
\end{equation}
This is warped AdS$_4$, not AdS$_3$. We set $\xi=0$ to avoid this enhancement. Using this condition we find the constraints \eqref{one}-\eqref{four} become
\begin{subequations}
\label{constraints}
\begin{align}
& \xi =0, \label{one1} \\[2mm]
& (\chi^{- \dagger} \chi^+ + \chi^{+ \dagger} \chi^-) = 0, \label{two1} \\[2mm]
& ||\chi^{\pm}||^2 = e^A c^{\pm}, \label{four1}
\end{align}
\end{subequations}
where neither of $c_{\pm}$ can be set to zero without setting one of  $\chi^{\pm}=0$. As we are intetested in ${\cal N}=(1,1)$ we will impose $c_{\pm} \neq 0$. We can then simply set them equal by rescaling $\zeta_{\pm}$, ie
\beq
c^{\pm}=c.
\eeq
We define the following forms
\begin{subequations}
\label{definitionsofSU(3)forms}
\begin{align}
& \chi^{\pm \dagger} \hat{\gamma} \chi^{\pm} = c e^A \cos \alpha, \\[2mm]
& \pm \chi^{\pm \dagger} \gamma_a \chi^{\pm} e^a = c e^A \sin \alpha V, \\[2mm]
& \chi^{ \pm \dagger} \gamma_a \chi^{\mp} e^a = c e^A \sin \alpha U, \\[2mm]
& \mp \frac{1}{2!} \chi^{\pm \dagger} \gamma_{ab} \hat{\gamma} \chi^{\mp} e^{ab} = c e^A \left( J + \cos \alpha U \wedge V\right), \\[2mm]
&- \frac{1}{3!} \chi^{\pm \dagger} \gamma_{abc} \hat{\gamma} \chi^{\pm} e^{abc} = c e^A \sin \alpha \left( \text{Im} \Omega \mp J \wedge U \right), \\[2mm]
& \pm \frac{1}{3!} \chi^{\mp \dagger} \gamma_{abc} \chi^{\pm} e^{abc} = c e^A \sin \alpha \text{Re} \Omega.
\end{align}
\end{subequations}
where $(J,\Omega,U,V)$ span an SU(3)-structure in d=8.
We report how the 1-form $K$ becomes, after imposing \eqref{constraints}, and the 2- and 5-form
\begin{subequations}
\begin{align}
K & = - c e^{2A} \left( v^+ + v^- \right), \label{definitionofK} \\[2mm]
\Xi & = c e^A \left(  e^A \sin \alpha (2 u \wedge U + (v_+ - v_- ) \wedge V ) + \cos\alpha  \frac{e^{2 A}}{f} u \wedge (v_+ - v_-) -2 f \left( J + \cos \alpha U \wedge V  \right)  \right), \label{definitionoftwoform} \\[2mm]
\Sigma & = 2 c e^A \bigg(- f \sin \alpha \text{Im}\Omega \wedge U \wedge V - f e^{3 A}\text{ vol(AdS$_3$)}\wedge ( U \wedge V + J \cos \alpha ) \nn \\[2mm]
& +\frac{ e^{2A}}{2 f}  \sin \alpha \left( u \wedge \left((v_+ + v_-)\wedge J \wedge U + (v_+ - v_- )\wedge \text{Im} \Omega \right) + v_+ \wedge v_- \wedge J \wedge V \right) \nn \\[2mm]
& + e^A \bigg( u \wedge (\text{Re} \Omega \wedge V + \cos \alpha \text{Im} \Omega \wedge U)-\bigg(\frac{1}{2} (v_+ - v_-)\wedge (\cos \alpha \wedge \text{Im} \Omega \wedge V - \text{Re} \Omega \wedge U)\bigg) \nn \\[2mm]
& +\frac{1}{4} (v_+ + v_-)\wedge J\wedge (J - 2 \cos \alpha  U \wedge V)\bigg)  \bigg). \label{definitionoffiveform}
\end{align}
\end{subequations}
Substituting these into \eqref{fluxconstraints} leads to a number of  necessary and sufficient $d=8$ conditions for supersymmetry, we quote in \eqref{BPS-conditionscomplex1}-\eqref{BPS-conditionscomplex7}. \eqref{definitionofK} implies that $K$ is clearly time-like, the equations \eqref{fluxconstraints} and the derived \eqref{BPS-conditionscomplex1}-\eqref{BPS-conditionscomplex7} are also sufficient now.

\section{S$^3$ spinors and bi-linears}\label{sec:S3}
In this appendix we construct the Killing spinors and the bilinears of S$^3$ we will use later. The symmetry group of the 3-sphere is SO(4) = SU(2)$_+ \times$ SU(2$)_-$.
We define the metric on S$^3$ as a foliation of S$^2$ over an interval as
\begin{align}
ds^2(\text{S}^3)= d\theta^2+ \sin^2 \theta \, ds^2(\text{S}^2),
\end{align}
where we define the 2-sphere in terms of embedding coordinates $y_i$ such that
\begin{align}
y_i^2 = 1, \qquad ds^2(\text{S$^2$}) = dy_i^2, \qquad \text{vol}(\text{S}^2)=\frac{1}{2}\epsilon_{ijk}y_i dy_j \wedge dy_k,
\end{align}
$y_i$ are triplets under the anti diagonal SU(2) one can form from SU(2)$_{\pm}$. On the 3-sphere one can define two sets of forms $ k^{\pm}_i$ that are respectively triplets with respect to SU(2)$_{\pm}$ and singlets with respect to SU(2)$_{\mp}$, they are defined in terms of $G\in\text{SU(2)}$ as
\begin{align}
\label{LeftandRightS3}
k^+_i = - i \text{Tr} \left( \sigma_i G^{-1} dG  \right), \qquad k^-_i = - i \text{Tr} \left( \sigma_i dG G^{-1}  \right),
\end{align}
such that they satisfy
\begin{align}
dk^{\pm}_i = \pm \frac{1}{2} \epsilon_{ijk} k^{\pm}_{j} \wedge k^{\pm}_{k}, \qquad ds^2(\text{S$^3$}) = \frac{1}{4} k^{\pm}_i k^{\pm}_i.
\end{align}
A convenient choice of $G$ for our parameterisation of the 3-sphere is
\begin{align}
G = e^{i \theta y_i \sigma^i}.
\end{align}
We define the SU(2)$_{\pm}$ Killing vectors in terms of  $k^{\pm}_i$
\begin{align}
K^{\pm }_i = \frac{1}{4} g^{ab} k^{\pm}_{ib} \partial_a,
\end{align}
where $g_{ab}$ is the metric of the 3-sphere and we denoted with $a,b,c$ its indices. \\
~\\
The Killing spinor equation of S$^3$ is
\beq
\label{KillingS3}
\nabla_{a} \xi^{\pm} = \pm \frac{i}{2}\gamma_{a} \xi^{\pm},
\eeq
where $\gamma_{a}$ are the Pauli matrices $\sigma_1,\sigma_2,\sigma_3$. These spinors are SU(2)$_{\pm}$ doublets and singlets with respect to SU(2)$_{\mp}$.  To show this claim we define the following SU(2)$_{\pm}$ doublets
\begin{align}
\xi^{\mathfrak{a}\pm} = \begin{pmatrix}
\xi^{\pm} \\
\xi^{\pm c} \\
\end{pmatrix}^{\mathfrak{a}},
\end{align}
where $\xi^{\pm c} = B_3 \xi^{\pm}$ where $B_3 = \sigma_2$. The Lie derivative along the Killing vectors acting on the bispinor gives rise to
\begin{align}
\mathcal{L}_{K^{\pm}_i} \xi^{\mathfrak{a}}_{\mp} = 0, \qquad \mathcal{L}_{K^{\pm}_i} \xi^{\mathfrak{a}}_{\pm} = \mp \frac{i}{2} \sigma_i \xi^{\mathfrak{a}}_{\pm}.
\end{align}
We define $\Xi_k^{st}$ whose components are defined as
\begin{align}
(\Xi^{st}_k)^{\mathfrak{ab}}= \frac{1}{k!} \xi^{\mathfrak{b}s\dag}\gamma_{n_1..n_k}\xi^{\mathfrak{a}t} e^{n_1... n_k},
\end{align}
where $t,s=\pm $ and $\mathfrak{a},\mathfrak{b}=1,2$. Here we write the $p$-forms for $p=0,1,2,3$
\begin{align}
& \Xi_0^{\pm \pm} = \mathbbm{1}_2, \qquad \Xi_1^{\pm \pm} = - \frac{1}{2} k^{\pm}_i \sigma^i, \qquad \Xi_2^{\pm \pm} = \mp \frac{i}{4} d k^{\pm}_i \sigma^i, \qquad \Xi_3^{\pm \pm} =  i \text{vol}(\text{S}^3)  \mathbbm{1}_2, \nn \\[2mm] 
& \Xi_0^{\pm \mp} = \cos\theta \mathbbm{1}_2 \pm i \sin\theta \sigma_i y^i, \qquad \Xi_1^{\pm \mp} = \pm i \left( \cos\theta \mathbbm{1}_2 \pm i \sin\theta \sigma^i y_i \right), \nn \\[2mm] 
& \Xi_2^{\pm \mp} = \sin^2 \theta \left( \pm \sin \theta \mathbbm{1}_2 -i \cos \theta \sigma^i y_i  \right) \text{vol}(\text{S}^2) -i \sin \theta \sigma^i dy_i \lrcorner \text{vol}(\text{S}^3) ,  \nn \\[2mm] 
& \Xi_3^{\pm \mp} = \left( i \cos \theta \mathbbm{1}_2 \mp \sin \theta \sigma^i y_i \right) \text{vol}(\text{S}^3),
\end{align}
such that
\begin{subequations}
\begin{align}
& d \left( \sin^3 \theta \, \text{vol}(\text{S}^2) \right) = 3 \cos \theta \, \text{vol}(\text{S}^3), \\[2mm]
&  d\left( \sin \theta\left( \cos \theta \sin \theta y_i \text{vol}(\text{S}^2) +  d \theta \wedge K_i \right) \right) = - 3 \sin \theta y_i \text{vol}(\text{S}^3).
\end{align}
\end{subequations}
Where $K_i$ are the three one-form charge under the antydiagonal SO(3),
\begin{align}
k^+_i -k^-_i = 4 \sin^2 \theta K_i, \qquad K_i = \epsilon_{ijk} y_jdy_k.
\end{align}
We choose a particular parameterization where the embedding coordinate $y_i$ of S$^2$ on S$^3$ are
\begin{align}
y_i=\left( - \cos \psi \sin \phi, - \sin \psi \sin \phi ,\cos \phi\right).
\end{align}
Using these coordinates, the Killing spinors that satisfy \eqref{KillingS3} are
\begin{align}
\xi^{\pm} = e^{\pm \frac{i}{2} \theta \sigma_1} e^{ \frac{1}{2} \phi \sigma_1 \sigma_2} e^{ \frac{i}{2} \psi \sigma_1 } \xi_0,
\end{align}
where $ \tilde{\xi}$ is a constant spinor, that in our parameterisation we choose to be
\begin{align}
\xi_0 = \begin{pmatrix}
- \frac{1}{\sqrt{2}} \\
  \frac{1}{\sqrt{2}} \\
\end{pmatrix},
\end{align}
where we choose the convention such that $|\xi^{\pm}|=1$.

\end{document}